%% file: main_preprint.tex
\documentclass[]{article}
\input{body/commands}
\usepackage{authblk}
\pdfoutput=1
\begin{document}
\title{Flexible Amortized Variational Inference in qBOLD MRI}
\author[1]{Ivor J.A. Simpson\footnote{Corresponding author: I.Simpson@sussex.ac.uk}}
\affil[1]{Department of Informatics, University of Sussex, Brighton, UK}
\author[1]{Ashley McManamon}
\author[2]{Bal{\'a}zs {\"O}rzsik}
\author[3]{Alan J. Stone}
\author[4]{Nicholas P. Blockley}
\author[2]{Iris Asllani}
\author[2]{Alessandro Colasanti}
\author[5]{Mara Cercignani}
\affil[2]{CISC, Brighton and Sussex Medical School, University of Sussex, Brighton, UK}
\affil[3]{Department of Medical Physics and Clinical Engineering, St. Vincent's University Hospital, Dublin, Ireland}
\affil[4]{School of Life Sciences, University of Nottingham, Nottingham, UK}

\affil[5]{CUBRIC, Cardiff University, Cardiff, UK}
%

\date{}

\maketitle

\input{body/abstract}


\input{body/body}
\bibliographystyle{plainnat}
\bibliography{references}
\newpage
\input{body/appendix}
\end{document}

%% file: body/commands.tex
\usepackage[]{geometry}
\usepackage[utf8]{inputenc}
\usepackage{amsmath, cases}
\usepackage{natbib}
\usepackage{xcolor}
\usepackage{graphicx}
\usepackage{bm}
\usepackage{caption}
\usepackage{subcaption}
\usepackage{hyperref}
\usepackage{url}

\usepackage{amssymb}

\newcommand{\oef}{OEF}
\newcommand{\dbv}{DBV}
\newcommand{\oefs}{\mathrm{OEF}}
\newcommand{\dbvs}{\zeta}
\newcommand{\ti}{\tau}
\newcommand{\sig}{S_t}
\newcommand{\dchi}{\Delta\chi_{0}}
\newcommand{\domega}{\delta\omega}
\newcommand{\rp}{R_2^\prime}
\newcommand{\y}{\textbf{y}}
\newcommand{\x}{\textbf{x}}
\newcommand{\X}{\textbf{X}}
\newcommand{\sstar}{\textbf{S}^*}
\newcommand{\M}{M}
\newcommand{\E}{\mathbb{E}}
\newcommand{\elbo}{\mathcal{L}}
\newcommand{\ptparams}{\theta}
\newcommand{\mub}{\bm{\mu}}
\newcommand{\Sigmab}{\bm{\Sigma}}
\newcommand{\mupt}{\mub_l(\sstar; \ptparams)}
\newcommand{\sigmapt}{\Sigmab_l(\sstar; \ptparams)}
\newcommand{\tauvec}{\bm{\tau}}
\newcommand{\stau}{S_{\tauvec}}
\newcommand{\truncnorm}{T\mathcal{N}}

\newcommand{\logitn}{L\mathcal{N}}

\newcommand{\ftparams}{\mathbf{\psi}}
\newcommand{\sigmafft}{\Sigmab_l(\sstar; \ftparams)}
\newcommand{\sigmaimnonet}{\mathbf{\Sigma_{\mathrm{im}}}}
\newcommand{\sigmaim}{\mathbf{\Sigma_{\mathrm{im}}}(\x,\ftparams)}
\newcommand{\mufft}{\mub_l(\sstar; \ftparams)}
\newcommand{\hct}{Hct}

%% file: body/abstract.tex
\begin{abstract}
Streamlined qBOLD acquisitions enable experimentally straightforward observations of brain oxygen metabolism. $R_2^\prime$ maps are easily inferred; however, the Oxygen extraction fraction (OEF) and deoxygenated blood volume (DBV) are more ambiguously determined from the data. As such, existing inference methods tend to yield very noisy and underestimated OEF maps, while overestimating DBV.

This work describes a novel probabilistic machine learning approach that can infer plausible distributions of OEF and DBV. 
Initially, we create a model that produces informative voxelwise prior distribution based on synthetic training data. 
Contrary to prior work, we model the joint distribution of OEF and DBV through a scaled multivariate logit-Normal distribution, which enables the values to be constrained within a plausible range.
The prior distribution model is used to train an efficient amortized variational Bayesian inference model. This model learns to infer OEF and DBV by predicting real image data, with few training data required, using the signal equations as a forward model. 

We demonstrate that our approach enables the inference of smooth OEF and DBV maps, with a physiologically plausible distribution that can be adapted through specification of an informative prior distribution. Other benefits include model comparison (via the evidence lower bound) and uncertainty quantification for identifying image artefacts. Results are demonstrated on a small study comparing subjects undergoing hyperventilation and at rest. We illustrate that the proposed approach allows measurement of gray matter differences in OEF and DBV and enables voxelwise comparison between conditions, where we observe significant increases in OEF and $R_2^\prime$ during hyperventilation.
\end{abstract}

%% file: body/body.tex
\section{Introduction}
Quantitative blood oxygen level dependent (qBOLD) magnetic resonance imaging (MRI) provides a mechanism for creating in-vivo maps of a parameter of brain oxygen metabolism, Oxygen extraction Fraction (OEF)~\citep{Yablonskiy1994}. According to the Fick’s principle~\citep{kety1948effects}, OEF relates the rate of brain oxygen metabolism (CMRO2) to that of oxygen delivery (corresponding to cerebral blood flow). This relationship assumes that the primary function of blood flow is the delivery of oxygen to the brain, and that brain metabolism drives the extraction of oxygen from blood into tissues.

Measurement of OEF may find application in the study of healthy brain physiology, as well as in several clinical applications~\citep{dunn2021impact} such as stroke~\citep{stone2019prospects,an2021mitochondrial}, dementia~\citep{correia2021oxygen}, MS~\citep{trapp2009virtual}, Parkinson’s Disease~\citep{henchcliffe2008mitochondrial}, Bipolar Disorder~\citep{andreazza2010mitochondrial}\citep{pinna2021neurometabolic} and Schizophrenia~\citep{prabakaran2004mitochondrial}, in which alterations in brain oxygen supply or utilization  play a prominent pathophysiological role.

The qBOLD model describes the relationship between OEF and the deoxygenated blood volume (DBV) through the dependency of the transverse relaxation rate, $\rp$ where $\rp = R_2^* - R_2$, on the level of oxygenation in the blood. 
Recent work by Stone and Blockley~\citep{stone2017streamlined} demonstrated the potential for a streamlined quantative BOLD acquisition that makes use of an asymmetric spin echo sequence \citep{wismer1988susceptibility}. 

\begin{figure*}[!hbt]
    \centering
    \includegraphics[width=0.7\textwidth]{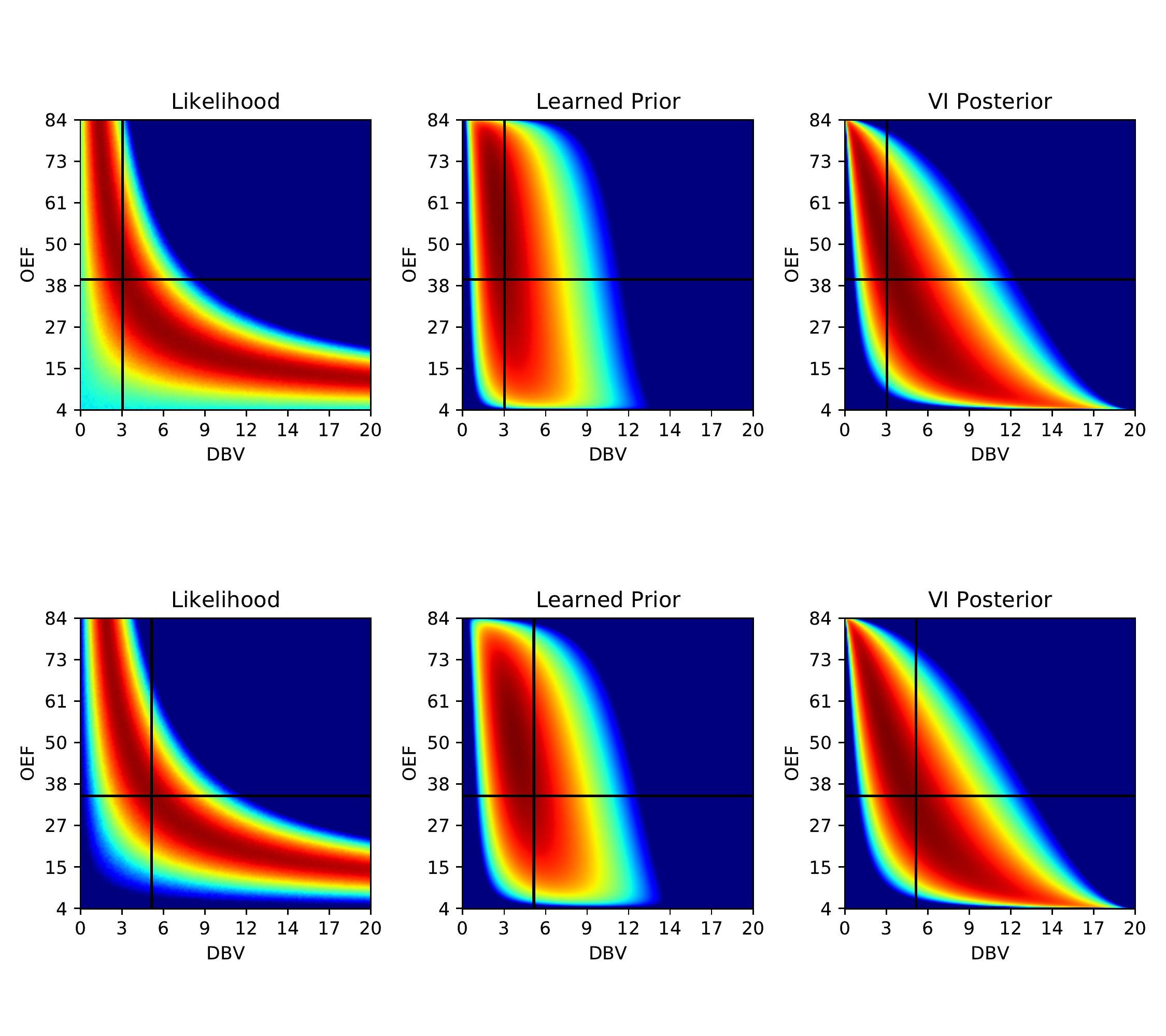}
    \caption{Illustration of log probability of OEF/DBV given simulated data, where the true values are given at the intersection of the black lines. Note that the likelihood (left column) supports a wide range of similarly likely parameters, some of which are physiologically unlikely. The middle column shows the result of our adaptive prior, trained using synthetic data. The right column illustrates our inferred approximate posterior distribution.}
    \label{fig:likelihood}
\end{figure*}

Performing maximum likelihood (weighted least squares) inference on the streamlined qBOLD data provides smooth and robust maps of $\rp$.  However, the inferred maps of OEF tend to be extremely noisy. This noisiness arises from a broad joint distribution of probable values for these parameters, given the data~\citep{sohlin2011susceptibility, christen2014mr, Cherukara2019}. This is illustrated in the left column of Figure \ref{fig:likelihood}. Accordingly, treating voxels as independent and taking a maximum likelihood approach to model fitting yields inconsistent results, even with high signal-to-noise ratio. Recently, a variational Bayes approach \citep{Cherukara2019} was proposed for this form of data, which yields some improvements in terms of enabling smoother predictions and objective model comparison. However, there are several limitations with this approach including: systematic under/over-estimation of OEF and DBV respectively; an approximated form of the true forward model is used, where the accuracy of the approximation varies for different parameters values; the approach requires sequential updates from an initial posterior distribution, and may yield inconsistent results for different initialisations; finally, this previous work describes the joint distribution of \dbv\ and $\rp$, which does not allow prior knowledge to be directly expressed for \oef. 

This paper contributes a flexible framework for efficient amortized variational Bayesian inference of voxelwise phyisological parameters from neuroimaging data with the following properties:
\begin{enumerate}
    \item Synthetic data is used to pre-train a convolutional encoder, from which informative voxelwise prior distributions are derived prior to observing any real data. These adaptive priors and our variational posteriors are illustrated in Figure \ref{fig:likelihood}. We demonstrate that our approach leads to interpretable inference results.
    \item Capable of using any differentiable non-linear forward models, without linearised mathematical approximations. We demonstrate an improvement compared to asymptotic approximations of the signal equations.
    \item Inference is performed using a feed-forward convolutional neural network, so has no initialisation dependence at inference time and is computationally efficient.
    \item Allows for flexible distribution choices for the parameters of interest; in this case we use logit-Normal distribution for OEF and DBV to restrict to physically possible values.
    \item Additional loss functions or regularisers can be trivially incorporated. In this work we experiment with a total-variation smoothness regulariser. 
\end{enumerate}
Furthermore, an open-source implementation and pre-trained models in TensorFlow are made available through GitHub\footnote{https://github.com/wearepal/qBOLD-VI}. We demonstrate our approach on streamlined asymmetric spin echo quantitative BOLD data, but it could be easily adapted for other qBOLD acquisitions or similar problems such as measuring cerebral blood flow using arterial spin labelling data.

\section{Background}
\subsection{Forward Model}
The theoretical qBOLD model of \citep{Yablonskiy1994}, describes the expected transverse relaxation signal profile given some assumptions regarding the tissue \citep{an2003impact}. The transverse signal, $S_t$, is described as an exponential function of the time $t$:
\begin{equation}
    \sig(t) = \exp(-R_{2}^{t} \cdot t) \cdot \exp\Bigg( -\dbvs \cdot \int_{0}^{1} \dfrac{(2+u)\sqrt{1-u}}{3 \cdot u^{2}} \Bigg(1 - J_{0} \Bigg(\dfrac{3}{2}\delta\omega t u \Bigg) \Bigg) du \Bigg).\label{eq:full_fwd_orig}
\end{equation}
where $\dbvs$ is \dbv\,  $J_0$ is the order zero Bessel function. $\domega$ is the characteristic frequency, given by:
\begin{equation}
    \domega = \gamma \cdot \frac{4}{3} \cdot \pi \cdot \dchi \cdot \hct \cdot \oef \cdot B_{0}
\end{equation}
where $\gamma$ is the gyromagnetic ratio of protons, $\dchi$ is the difference in magnetic susceptibility between entirely oxygenated, and entirely deoxygenated red blood cells, $\hct$ is is the blood haematocrit (volume \% of red blood cells) , $\oef$ is the oxygen extraction fraction, and $B_{0}$ is the strength of the main static magnetic field. 

The asymmetric spin echo (ASE) qBOLD sequence~\citep{stone2017streamlined} makes uses of a refocusing pulse at time $(TE-\ti) / 2$, leading to an effective echo time of $TE-\ti$.  The signal is read-out at constant time TE across all the echoes, and we can now write the signal as a function of $\ti$:
\begin{equation}
    \sig(\ti) = \exp(-R_{2}^{t} \cdot TE) \cdot \exp\Bigg( -\dbvs \cdot \int_{0}^{1} \dfrac{(2+u)\sqrt{1-u}}{3 \cdot u^{2}} \Bigg(1 - J_{0} \Bigg(\dfrac{3}{2}\delta\omega \ti u \Bigg) \Bigg) du \Bigg). \label{eq:full_fwd}
\end{equation}
$\ti$ is varied through multiple acquisition to sample the $\rp$ decay. Model parameter symbols and constants are defined in table \ref{tab:params}.

\begin{table}
\begin{center}
\begin{tabular}{ |c|c|c| } 
 \hline
 \textbf{Symbols} & \textbf{Meaning} & \textbf{Constant/Inferred}  \\ \hline \hline
 $\dbvs$  & Deoxygenated blood volume (\dbv) & Inferred \\ \hline
$\oefs$ & Oxygen extraction fraction (\oef) & Inferred \\ \hline 
$\rp$ & Reversible transverse relaxation rate & Indirectly inferred \\ \hline\hline
$\hct$ & Fractional haematocrit & 0.34 \citep{He2007}  \\ \hline
$\dchi$ & Magnetic Susceptibility Difference & 2.64e-7 \citep{He2007} \\ \hline
$\gamma$ & Gyromagnetic ratio of protons & 2.675e8 \\ \hline
$R_2^t$ & R$_2$ decay of brain tissue & $11.5$ \citep{He2007} \\
 \hline
\end{tabular}
\end{center}
\caption{Key model parameters}
\label{tab:params}
\end{table}

\subsection{Asymptotic Approximations}
Prior works~\citep{Yablonskiy1994, an2003impact, stone2017streamlined} have decomposed this forward model into two asymptotic mono-exponential equations for different regimes dependent on a characteristic time $t_c$.
\begin{equation}
    \sig(\tau) = \begin{cases}
    S_{0} \cdot \exp(-R_{2}^{t} \cdot TE) \cdot \exp \left (-\dfrac{3}{10} \cdot \dfrac{(R_{2}' \cdot \tau)^{2}}{\dbvs} \right ) & \mid \tau \mid < t_{c} \\
    S_{0} \cdot \exp(-R_{2}^{t} \cdot TE) \cdot \exp(\dbvs - R_{2}' \cdot \tau) & \mid \tau \mid > t_{c}.
    \end{cases}
\end{equation}
The transition points between these regimes are imperfectly characterised. \citep{stone2017streamlined} indicate that $t_c$ should be at $\frac{1.5}{\delta \omega}$, whereas \citep{lee2018interleaved} opt for $\frac{1.0}{\delta \omega}$, \citep{Cherukara2019} advocated a process of adapting $t_c$ to best match equation \ref{eq:full_fwd}. Although these asymptotic forms provide for a straightforward numerical solution, the issues regarding the choice of regime demonstrate the benefits of using the full forward model where possible. This work can encapsulate either the asymptotic formulation or the full forward model.





\subsection{Two Compartment Model}
Several approaches have been proposed to model the signal contribution from venous blood. Similarly to \citep{Cherukara2019}, we adopt the intravascular compartment model of \citep{berman2018transverse} to describe a venous blood compartment:
\begin{multline}
        S^{b}(\tau) = \exp(-R_{2}^{b} \cdot TE) \cdot \exp \Bigg( -\dfrac{\gamma^{2}}{2} G_{0} t_{D}^{2} \cdot \Bigg( \dfrac{TE}{t_{D}} + \sqrt{\dfrac{1}{4} + \dfrac{TE}{t_{D}}} + \dfrac{3}{2} \\
        - 2\sqrt{\dfrac{1}{2} + \dfrac{TE + \tau}{t_{D}}} - 2\sqrt{\dfrac{1}{4} + \dfrac{TE - \tau}{t_{D}}} \Bigg) \Bigg)
        \label{eq:blood}
\end{multline}
where the characteristic diffusion time, $t_{D} = r_{b}^{2}/D_b$ with $r_b = 2.6\mu m$ and $D_b=2\mu m^2ms^{-1}$ corresponding to the size and diffusion rate of red blood cells, respecitvely \citep{berman2018transverse, Cherukara2019}. $G_{0}$, the mean square field inhomogeneity is calculated using equation from \citep{berman2018transverse}:
\begin{equation}
     G_0 = \dfrac{4}{45}\hct(1 - \hct)(\dchi B_{0})^{2}
\end{equation}

The total signal is a combination of the signal from Eq. \ref{eq:full_fwd} and Eq. \ref{eq:blood} weighted by the apparent deoxygenated blood volume, $\dbvs'$. 
\begin{equation}
    S^{total} = (\dbvs'S^{b}(\tau) + (1-\dbvs')S(\tau))
\end{equation}
where $n_b$ is the relative spin density of blood (0.775) \citep{Ernst1993AbsoluteWater}, and $\dbvs' = m_b \cdot n_{b} \cdot \dbvs$. $m_b$ is the steady-state magnetisation of blood, which for this sequence is given by \citep{Cherukara2019thesis}:
\begin{equation}
    m_b = 1-\left(2-\exp\left(-\frac{TR-TI}{T1_b}\right)\right)\exp\left(-\frac{TI}{T1_b}\right)
\end{equation} 
where $TR=3s$, $TI=1.21s$ is the inversion time for the FLAIR pulse~\citep{stone2017streamlined} and $T1_b=1.58s$ is the blood T1 relaxation time~\citep{Cherukara2019thesis}.



\subsection{Variational Inference}
\subsubsection{Theory}
Bayesian inference provides a principled and coherent method for inferring the parameters of a model from data, given some prior knowledge. Bayes' theorem states:
\begin{equation}
    p(\theta | \x, \M) = \frac{p(\x | \theta, \M) p(\theta)}{p(\x | \M)}
\end{equation}
where $\theta$ are the parameters of model $\M$ and $\x$ is data. The model evidence, $p(\x | \M)$, is intractable for most problems. This complicates inference of the posterior over model parameters $p(\theta | \x, \M)$.

Variational Bayesian Inference (VI) provides a tractable approach to estimating approximate parameter distributions. The fundamental idea is to approximate the posterior of the model parameters, $p(\theta|\x)$, using an analytic distribution, $q(\theta | \x)$. Commonly $q(\theta | \x)$ is chosen as a multivariate normal, $p(\theta|\x) \approx q(\theta|\x) = \mathcal{N}(\mu_\theta, \Sigma_\theta) $. The parameters of $q(\theta| \x)$ are chosen by maximising the log evidence lower-bound (ELBO), $\elbo$:
\begin{eqnarray}
\log p(\x | \M) = \log \int p(\x, \theta | \M) d\theta &\geq& \E_{q(\theta|\x)} \left[\log \frac{p(\x|\theta, \M) p(\theta)}{q(\theta|\x)}\right]  \\
&\geq& \E_{q(\theta|\x)}[\log(p(\x| \theta, \M)] - D_{KL}(q(\theta | \x)||p(\theta)) = \elbo
\end{eqnarray}
where $\E$ corresponds to the expectation of the bracketed quantity with respect to the subscript. Detailed derivations of variational inference are widely available and we refer readers to consult~\citep{blei2017variational}.
\subsubsection{Model evidence}\label{sec:model_evidence}
The model evidence, sometimes called the marginal likelihood, $p(\x | \M) = \int p(\x| \theta, \M)p(\theta)d\theta$, provides a summary of how well the model explains the data. As the ELBO lower-bounds the log model evidence, it has been used to objectively compare choices of models to explain data~\citep{beal2003variational}. This has been shown to be particularly useful in cases such as neuroimaging, where the ``true" parameters are unobservable and the evidence is computationally infeasible, e.g. \citep{chappell2008variational, friston2007variational}. 

\subsubsection{Inference} 
Traditional VI was performed through a variant of expectation-maximisation~\citep{beal2003variational}, making use of conjugacy to allow analytic updates without the need for sampling. \citep{chappell2008variational} demonstrated an approach for using VI with non-linear forward models using a 1st order Taylor series. However, these approaches require initial estimates of posterior distributions and are susceptible to local minima~\citep{blei2017variational}.

The Variational Auto Encoder (VAE) \citep{kingma2013auto, rezende2014stochastic} demonstrated how neural networks can learn to perform variational inference, amortized over a training set $\X$. This is achieved by creating a neural network to parameterise the approximate distribution, $q_\Theta(\theta | \x)$, where $\Theta$ are the parameters of the neural network. This network can be trained through backpropagation by approximating the expectation $\E_{q(\theta|x)}\left[ p(\x | \theta) \right]$ using samples from $q_{\Theta}(\theta|\x)$. Leading to a loss of:
\begin{equation}
    -\elbo \approx \frac{1}{N}\sum_i^N \left[ D_{KL}(q_\Theta(\theta|\x^i)||p(\theta)) - \frac{1}{N_s}\sum_j^{N_s}\log(p(\x^i|\theta^j, M)) \right]
\end{equation}
where $N$ is the number of data samples and $N_s$ is the number of samples from the approximate posterior, $\theta^j \sim q_\Theta(\theta|\x^i)$, used to evaluate the log-likelihood.

\section{Methods}
\subsection{Overview}
\begin{figure}[!hbt]
    \centering
    \begin{subfigure}[b]{0.5\textwidth}
         \centering
        \includegraphics[width=\textwidth]{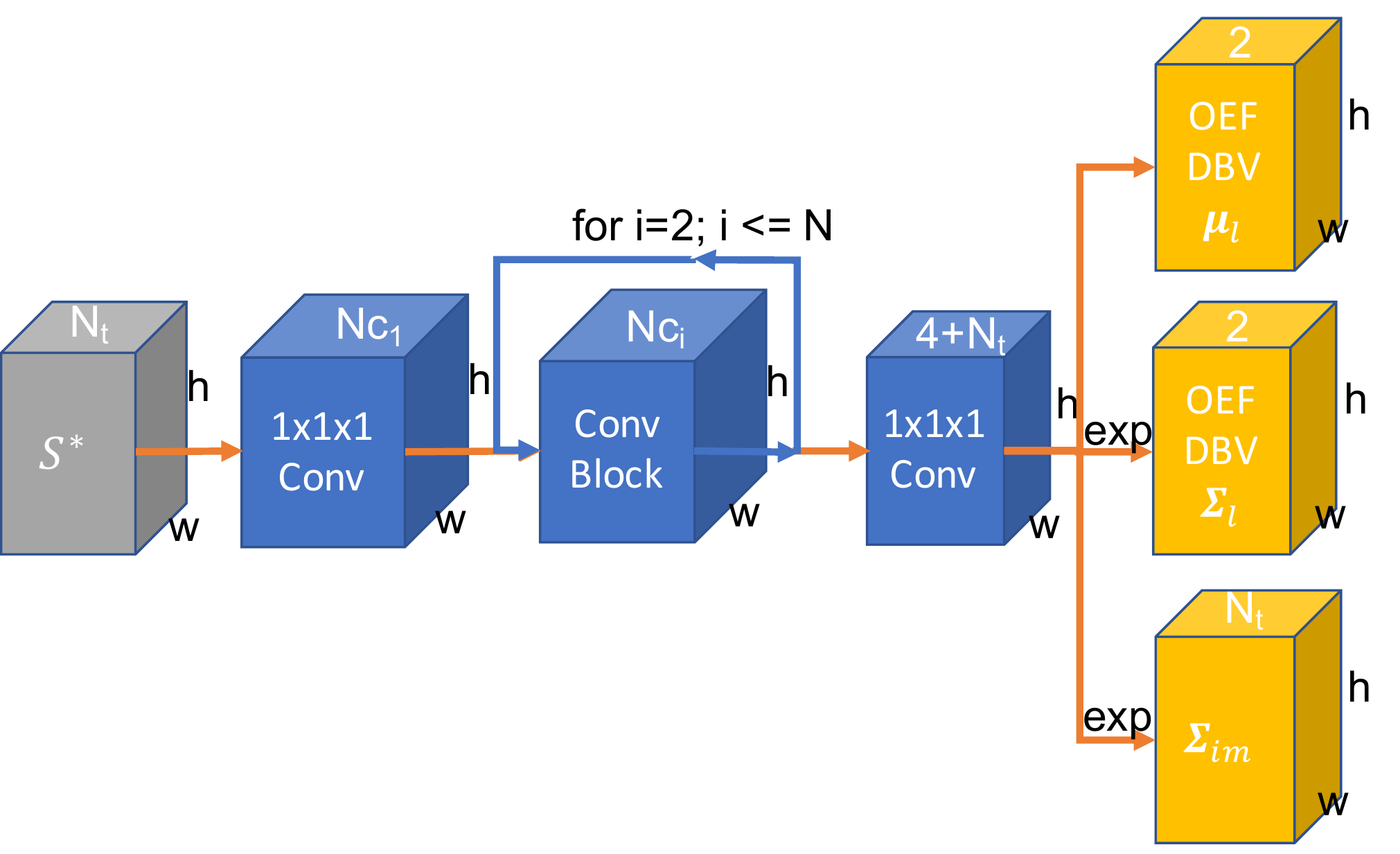}
        \caption{}
        \label{fig:encoder}
    \end{subfigure} \\ 
    \begin{subfigure}[b]{0.275\textwidth}
    \centering
        \includegraphics[width=\textwidth]{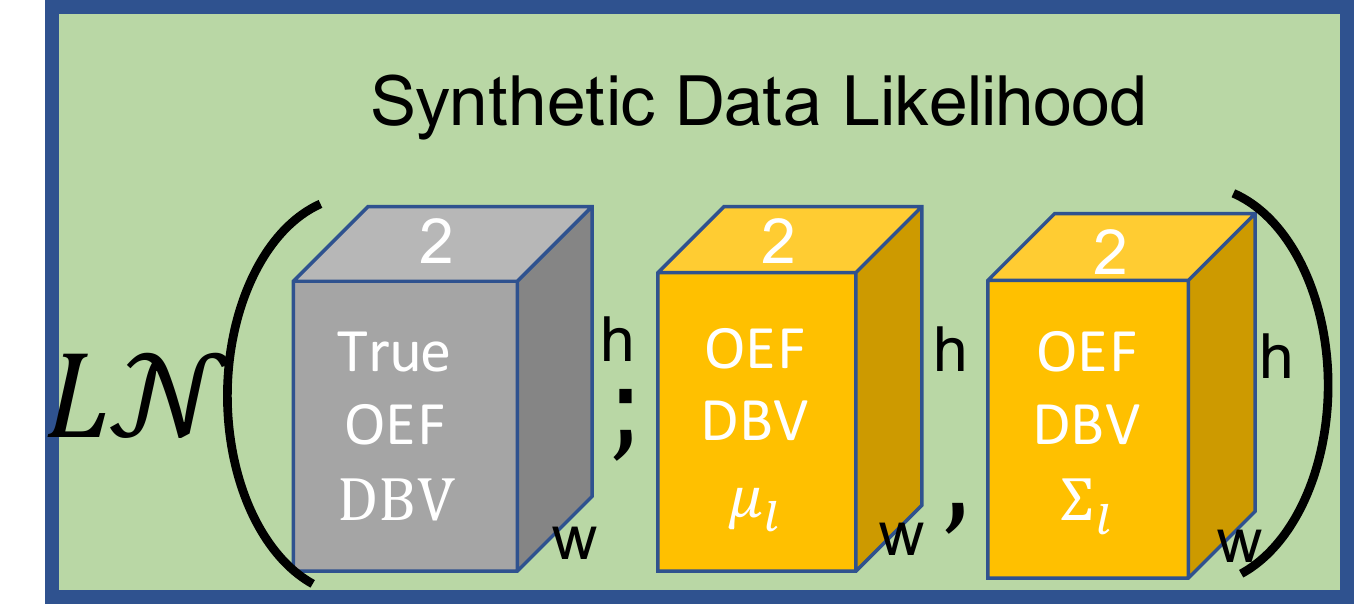}
        \caption{}
        \label{fig:synth_likelihood}
    \end{subfigure}
    \begin{subfigure}[b]{0.38\textwidth}
    \centering
        \includegraphics[width=\textwidth]{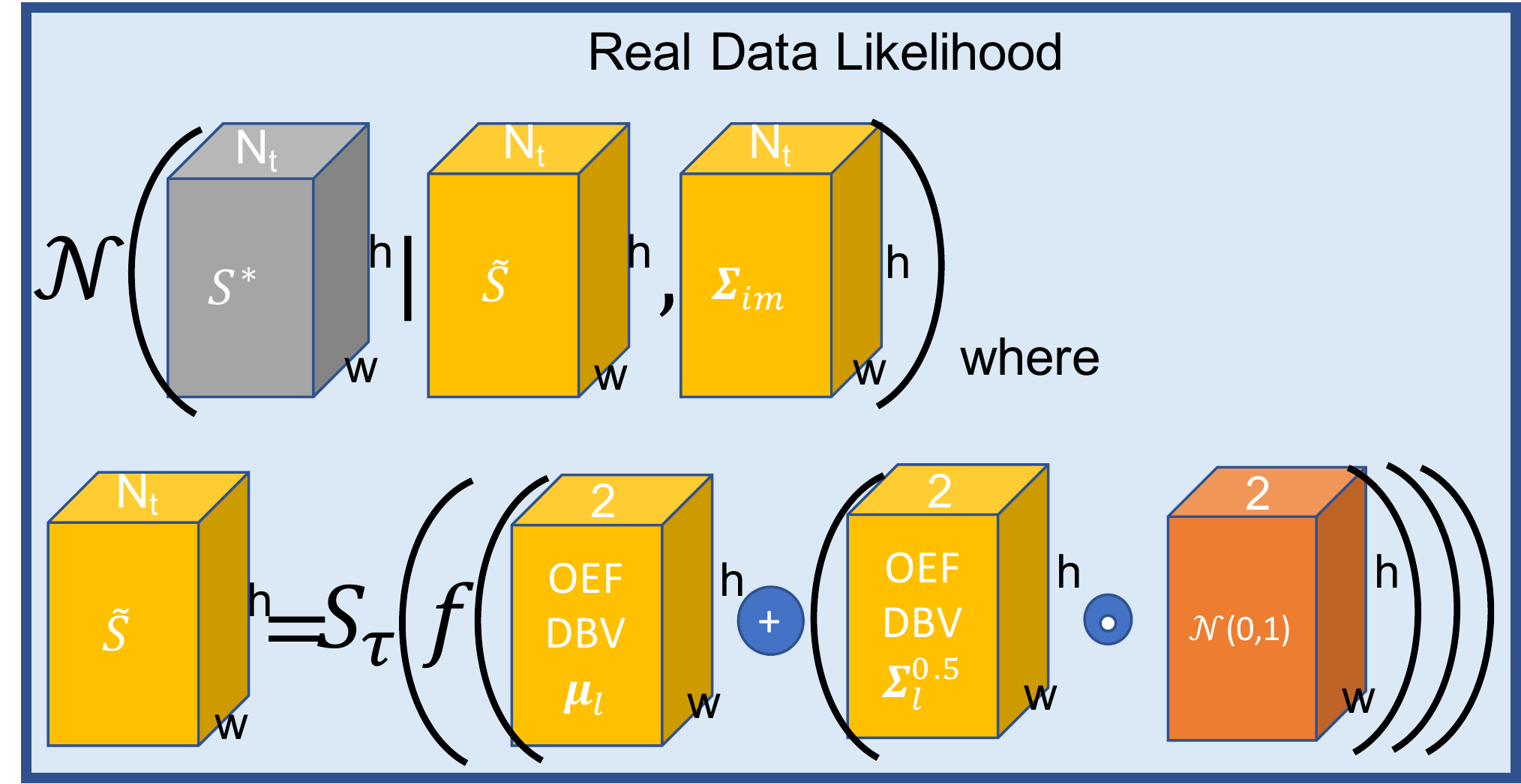}
        \caption{}
        \label{fig:real_likelihood}
    \end{subfigure}
    \caption{a) shows the overall encoder architecture, where the specifics of the convolutional blocks are described in more detail in Figure \ref{fig:conv_block}. Tensors are shown for 2D ($h \times w \times$ channels), rather than 3D images, for simplicity of presentation. Grey blocks correspond to observations, where $S^*$ is the normalised input data, orange blocks are model outputs and blue are intermediate representations. Note these diagrams illustrate independent, rather than correlated \oef\ and \dbv\ distributions.
    A block diagram to illustrate the synthetic data likelihood is given in \ref{fig:synth_likelihood}, where the true values are known and are logit-Normal distributed (as described in section \ref{sec:param_dist}). 
    The real data likelihood is illustrated in Figure \ref{fig:real_likelihood}, where the reparameterisation trick followed by a scaled and shifted logit transformation, $f$, is used to generate samples of $\dbv$ and $\oef$ from which a simulated signal is synthesised $\hat{S}$. The likelihood model is a Normal distribution, with heteroscedastic noise parameters given by $\sigmaimnonet$ \label{fig:overview}.} 
\end{figure}

Our approach is summarised in Figure \ref{fig:overview}. The model consists of a convolutional neural network, referred to as an encoder. The encoder takes an image of dimensions $h \times w \times d \times N_t$, where $N_t$ is the size of $\tauvec$, and predicts a $h \times w \times d$ scaled set of logit-Normal distribution parameters for $\oef$ and $\dbv$. See section \ref{sec:param_dist} for details on the form of the distribution. The encoder also outputs a map $\Sigma$ of size $h \times w \times d \times N_t$, which contains the expected voxelwise variance for the predicted signal.

The inference model is trained in a two stage process using both synthetic and real data. We firstly synthesise a number of independent voxelwise signals with varying levels of noise. These synthetic data have known values for $\oef$ and $\dbv$ and accordingly this network can be trained in a supervised fashion. The likelihood for pre-training on synthetic data is illustrated in Figure \ref{fig:synth_likelihood}. At this stage, due to the spatial independence of the simulated signals, the encoder only uses $1\times1\times1$ convolutions and so is effectively a voxelwise multi-layered perceptron (MLP). 

Subsequently, the encoder is fine-tuned on real MRI data. In this phase we  optionally add additional larger sized convolutional components to the encoder, see section \ref{sec:arch} for details. Samples from the predicted $\oef$ and $\dbv$ distributions are fed into the forward model to reconstruct the data, as illustrated in Figure \ref{fig:real_likelihood}. In this variational training procedure, the voxelwise distributions predicted by the synthetically pre-trained model are treated as prior information. 

\subsection{Synthetic Data Generation}
We consider a similar process to \citep{domsch2018oxygen, hubertus2019using} of generating a set of synthetic data using known $\dbv$ and $\oef$ across the range of plausible values. The clean signal data is generated using the integral of equation \ref{eq:full_fwd}, which is solved numerically using Simpson's rule, where the domain is broken into 64 equal-width trapezoids.

We henceforth refer to the signal generation function as $\stau : \mathbb{R}^2 \mapsto  \mathbb{R}^{N_t}$, which maps from $\dbv$ and $\oef$ to a predicted set of $N_t$ signals with timings given by $\tauvec$.

To improve our robustness to noise, we simulate additive Gaussian noise applied to the clean signal. The SNR is known to vary given the value of $\tau$~\citep{stone2017streamlined}, with peak SNR at the spin echo image ($\tau=0$). We calculate a spin-echo normalised noise level from our training data, which was found to be consistent across examples, even if the noise variance itself was not. These normalised noise statistics were used to create additive noise at a range of spin-echo SNRs (between 50 and 120).



\subsubsection{Data Normalisation}
In order to make real and synthetic images comparable, they need to be normalised before being fed into the model. We normalise each voxel independently by dividing the observed signals by the spin-echo image, $\tau=0$. We subsequently log-transform the resulting values, mirroring previous works~\citep{stone2017streamlined}. We refer to the normalised data as $\sstar$. \label{sec:data_norm}

\subsection{Parameter distributions}\label{sec:param_dist}
We choose to model $\oef$ and $\dbv$ using a scaled logit-Normal ($\logitn$) distributions, as illustrated in $\ref{fig:likelihood}$. We define the distribution such that a random variable $\bm{\alpha} \sim \logitn(\mub_l, \Sigmab_l, \textbf{s}, \textbf{o})$ is given by $f(\bm{\beta}) = \textbf{s} L(\bm{\beta}) + \textbf{o}$ , where $L$ is the logistic function and $\bm{\beta} \sim \mathcal{N}(\mub_l, \Sigmab_l)$. Effectively this means the logits, inverse logistic values, of a shifted (by $\textbf{o}$) and scaled (by $\textbf{s}$) version of $\bm{\alpha}$ follow a multivariate Normal distribution. This has the benefit of constraining the probable domain of $\bm{\alpha}$ to being within a predefined range, while permitting correlations between dimensions. In our case, we constrain the random variables for $\oef$ and $\dbv$ parameters within a realistic range, and allow correlations in the logits. 

We define the probability density function for observed values $\y$ as:

\begin{equation}
    p(\hat{\y}; \mub_l, \Sigmab_l) = \frac{1}{\sqrt{2\pi}|\Sigmab_l^{\frac{1}{2}}|}\frac{1}{\sum_i^2 \hat{\y}_i(1-\hat{\y}_i)}\exp \left(-0.5(\mathrm{logit}(\hat{\y})-\mub_l)^T\Sigmab_l^{-1}(\mathrm{logit}(\hat{\y})-\mub_l)\right) \label{eq:logit_normal_pdf}
\end{equation}
where $\hat{\y} = (\y - \textbf{o})/\textbf{s}$, and $\mathrm{logit}$ is the inverse of the logistic function $\mathrm{logit}(\hat{\y}) = \log (\hat{\y}/(1-\hat{\y})$.

For our model, we use the following parameterisation:
\begin{equation}
    p(\Phi) = p(\oef, \dbvs) = \logitn(\mub_l, \Sigmab_l, \textbf{s}, \textbf{o}) 
\end{equation}
where the subscript $_l$ is used to distinguish these parameters from the mean and covariance of a multivariate Normal distribution. We use $p(\Phi)$ to denote the joint distribution of $\oef$ and $\dbv$, $\mub_l = \begin{bmatrix} \mub_{\oef}, \mub_{\dbvs}  \end{bmatrix}^T$, $\textbf{s} = \begin{bmatrix} 0.8,  0.3\end{bmatrix}^{T}$, $\textbf{o} = \begin{bmatrix} 0.05,  0.001\end{bmatrix}^{T}$. $\Sigmab_l$ is either a diagonal covariance matrix, or a lower-triangular Cholesky decomposition $\Sigmab_l = L_lL_l^T$ accounting for correlations in the logits between parameters. Note that equation \ref{eq:logit_normal_pdf} is not the standard definition of a multivariate logit-Normal distribution, in which the dimensions of x are considered exclusive observations (i.e. they sum to 1). Instead, we are modelling non-exclusive proportions where correlations are permitted.

In practice, when we use a non-diagonal $\Sigmab$ we whiten the residual term by the inverse transposed Cholesky matrix, $\begin{bmatrix}\mathrm{logit}(\y)-\mub \end{bmatrix}^\top L_l^{-\top}$ and take the sum of squares. Samples can be easily draw from this parameterisation by $f(\mub_l + L_l\mathbf{z})$, where $\mathbf{z}\sim \mathcal{N}(0,1)$. As described in Figure \ref{fig:overview} we infer these parameters for each voxel using a convolutional neural network, $\mupt$ and $\sigmapt$, where $\sstar$ is the normalised input data and $\ptparams$ corresponds to the network parameters.

\subsection{Distribution of synthetic parameters}
We discovered that the distribution of synthetic values for $\oef$ and $\dbv$ has a pronounced effect on the predictions from the model. Originally, we considered training using a uniform distribution in a valid range for each parameter. In such a scenario, we encountered the systematic underestimation of $\oef$ and $\dbv$ that was reported in prior work using similar data~\citep{stone2017streamlined, Cherukara2019}. We investigate using more informative synthetic distributions to introduce our expected prior distributions for $\oef$ and $\dbv$. Specifically, we find that distributions of $\oef = \mathcal{N}(40\%, 20\%)$ and $\dbv = \mathcal{N}(2.5\%, 2\%)$ provide relatively wide, reliable synthetic samples that alleviate under-estimation of $\oef$. The choice of this distribution is explored in section \ref{sec:prior_dist}.

\subsection{Synthetic Training}
During the synthetic training phase, the true OEF and DBV values, $\oef^*$ and $\dbv^*$ respectively, are known. As such, we can simply maximise the probability (eq \ref{eq:logit_normal_pdf}) of the true values given the synthetic signals.


\subsection{Learning Amortized Variational Inference from Real Data}
The previously described inference network is trained from synthetic data, where it  learns to predict the relevant biophysical parameters using simulated signals. However, it is likely that real observed MRI data will contain unforeseen variations from the simulated signals, such as partial voluming and other sources of signal. As such, the synthetically trained model will not provide an optimal explanation for real data. Furthermore, as the synthetic data is simulated as independent voxels we cannot learn to make spatially smooth predictions.

We now consider our inference networks as parameterising a variational distribution, $q(\Phi)$:
\begin{equation}
    p(\Phi | \sstar) \approx q(\Phi | \sstar) = :L\mathcal{N}(\mufft, \sigmafft, \textbf{o}, \textbf{s})
\end{equation}
where we extend our inference networks $\mufft$, $\sigmafft$ and refer to this parameterisation by $\psi$. $\psi$ contains additional parameters compared to $\theta$, where these additional parameters are randomly initialised.

In this VI approach, we choose to use predictions from a synthetic pre-trained model as informative prior information on the parameter logit distribution:  $p(\Phi) = L\mathcal{N}(\mupt, \sigmapt)$. We consider this prior knowledge as it arises from learning purely on simulated data. These informative priors provide a strong grounding for inference. 

As the true parameters $\oef$ and $\dbv$ are unobservable for real data, we instead simulate our observations using the forward model of the signal given our estimated parameters. We use a multivariate Normal likelihood function, with a diagonal covariance matrix $\sigmaim$ predicted as part of the model. Accordingly, the expected likelihood of the ELBO can be rewritten as:
\begin{eqnarray}
    E_{q(\Phi|\sstar)}\left[p(\sstar | \Phi, \M)\right] &=& \int \mathcal{N}(\sstar ; \stau(\Phi), \sigmaim, \M) q(\Phi|\sstar, \psi) \\
    &\approx& \sum_j^{N_s} \mathcal{N}(\sstar ; \stau(\Phi^j), \sigmaim))
\end{eqnarray}
where $\Phi^j \sim q(\Phi|\sstar)$. We can now write our approximate ELBO as:
\begin{eqnarray}
\mathcal{L} &=&  - E_{q(\Phi|\sstar)}\left[p(\sstar | \Phi, \M)\right] + D_{kl}(q(\Phi | \sstar) || p(\Phi | \sstar) \label{eq:elbo} \\
&\approx& - \sum_j^{N_s} \mathcal{N}(\sstar; \stau(\Phi^j), \sigmaim)) \\& & +  D_{KL}\left(L\mathcal{N}(\mufft, \sigmafft) || L\mathcal{N}(\mupt, \sigmapt)\right) \nonumber
\end{eqnarray}

where $D_{KL}$ is approximated using samples where a correlated logit-Normal distribution is used, and the implementation in TensorFlow probability~\citep{dillon2017tensorflow} is used for independent logit-Normal distributions for \oef\ and \dbv.


\subsubsection{Additional Regularisation}
One expected property in biological systems is spatial smoothness, i.e. adjacent voxels take similar values. This prior knowledge can help to further reduce the effects of noise compared to independent priors. Although principled approaches exist to infer spatial regularisation~\citep{groves2009combined}, these have not been demonstrated for multivariate logit-Normal distribution, and in any case add a higher level of complication to $p(\Phi)$. In this work we introduce spatial regularisation in a more ad-hoc way by including an additional total-variation loss on the transformed predicted means:
\begin{equation}
    TV(\hat{\mub}) = \frac{1}{(w-1)(h-1)d}\sum_{x=1}^{w-1} \sum_{y=1}^{h-1} \sum_{z=1}^d \left( |\hat{\mub}_{x,y,z} - \hat{\mub}_{x+1,y,z} | + |\hat{\mub}_{x,y,z} - \hat{\mub}_{x,y+1,z} | \right)
\end{equation}
where $\hat{\mub} = f(\mufft)$. Note that we only penalise spatial smoothness in the axial plane, due to the thick slices (and slice gap) in our data.

This gives us a total loss of:
\begin{equation}
    \mathcal{L}_{tv} = \mathcal{L} + \lambda TV(\mufft) \label{eq:loss}
\end{equation}
where $\lambda$ controls the magnitude of the smoothness regularisation.

\subsection{Network Architecture} \label{sec:arch}
The network trained on synthetic data considers voxels independently and consequently is best modelled using a multilayer perceptron (MLP), which can equivalently be considered as a 3D convolutional neural network (CNN) with a kernel size of 1 voxel. 

To incorporate both the pre-trained MLP, which considers voxels independently, and spatial information, we use a gated version of a residual blocks with larger (3x3x1) convolutional kernels~\citep{he2016deep}. Our convolutional block design is illustrated in the appendix in Figure \ref{fig:conv_block}. This residual block structure allow the network to encapsulate a wider receptive field, where the output is weighted by a learned gating value. For the gating function, we offset the logits by $-3$ such that the synthetically trained MLP is relied on initially when training the variational model.

\subsubsection{Training}
In our experiments, we opt for 2 MLP/gated-residual blocks, with 60 units throughout. 
During training, random $25 \times 25$ crops were taken of the data in x and y, and all the 8 slabs in z were used. The loss function was masked to only consider voxels in the brain region.

For training the model we used the AdamW optimiser~\citep{loshchilov2018decoupled}, with learning rate $2e^{-3}$ and weight decay rate $2e^{-4}$ for the synthetic data, and a learning rate of $5e^{-3}$ and a weight decay rate of $2e^{-4}$ for real data. We used batch sizes of 512 for synthetic data and 38 for real data. We found the use of stochastic weight averaging~\citep{izmailov2018averaging} while training with synthetic data improved the robustness of the overall training procedure. We trained for 1400 iterations on synthetic data, at which point the model converges, and 4000 iterations on real data. To ensure convergence on the real data, we linearly decay the learning and AdamW weight decay by a factor of 100 over the training period.

These parameters were chosen through a combination of parameter sweeps and empirical experimentation while monitoring training stability and the ELBO on validation data.

\section{Materials}

\subsection{Data}
Two datasets were used in this paper. The first was used for training the model consisting of 22 subjects, some of which had repeated scans, covering the inferior, superior or central regions of the brain. 38 separate scans were used from this dataset. The second dataset was used for validation, and we henceforth refer to as the study dataset. The study dataset is concerned with measuring changes in metabolism while undergoing hyperventilation. Hyperventilation is expected to have no effect on the cerebral metabolic rate of oxygen. However, OEF to expected to increase to compensate for the increased oxygen supply through the vasculature. This dataset consisted of 6 subjects, where data was acquired at rest and during hyperventilation. In the latter condition, participants were instructed to increase their ventilatory rate and depth, aiming for an end-tidal partial pressure of carbon dioxide (PETCO$_2$) of less than 20mmHg, assessed via capnograph, during MRI. In our analysis we have excluded one of the subjects whose hyperventilation scan suffered from artefacts, this image was identified as problematic based on the ELBO and predictive uncertainty measures, and visual inspection of the data confirmed the issue. We provide details of this in section \ref{sec:uncertainty}

These data were acquired with the FLAIR-GESEPI-ASE sequence described in \citep{stone2017streamlined}, with an in-plane axial resolution of 96x96, with 8 slices. The voxel dimensions were 2.3mm $\times$ 2.3mm $\times$ 5mm with a 2.5mm slice gap. For each acquisition, 11 values of $\tauvec$ were used, from -16ms to 64ms, with a spacing of 8ms. Both magnitude and phase images were saved, but only the magnitude data were examined in this work.

All data was motion corrected using the ``MCFLIRT"~\citep{jenkinson2001global} script from FSL~\citep{jenkinson2012fsl}, where the $\ti = 0$ image was chosen as the reference volume and sinc interpolation was used. Brain extraction was performed on the motion corrected ASE data using BET~\citep{smith2002fast}, where the average ASE volume was used to create the mask.

\subsection{Ethics statement and data availability}
Data acquisition was approved by the Brighton and Sussex Medical School Research and Governance Ethics Committee (REF 09/156/CER), and all participants provided written informed consent. 
The anonymised raw and preprocessed hyperventilation study dataset will be made freely available through Zeonodo.org on acceptance of this paper. All the code is freely available under an MIT license from \url{https://github.com/wearepal/qBOLD-VI}.

\section{Experiments}
The analysis of all of our results is performed on real data where we use the ELBO  (eq. \ref{eq:elbo}) of our validation dataset to compare model performance. ELBO increases with improved explanation of the original MR image data and closer accordance with the synthetically trained prior. We do not consider synthetic data useful for evaluation as it will not capture all sources of data variability and may prefer models that overfit to the synthetic data.

\subsection{Effect of smoothness loss}
\begin{figure}[!hbt]
    \centering
    \includegraphics[width=\columnwidth]{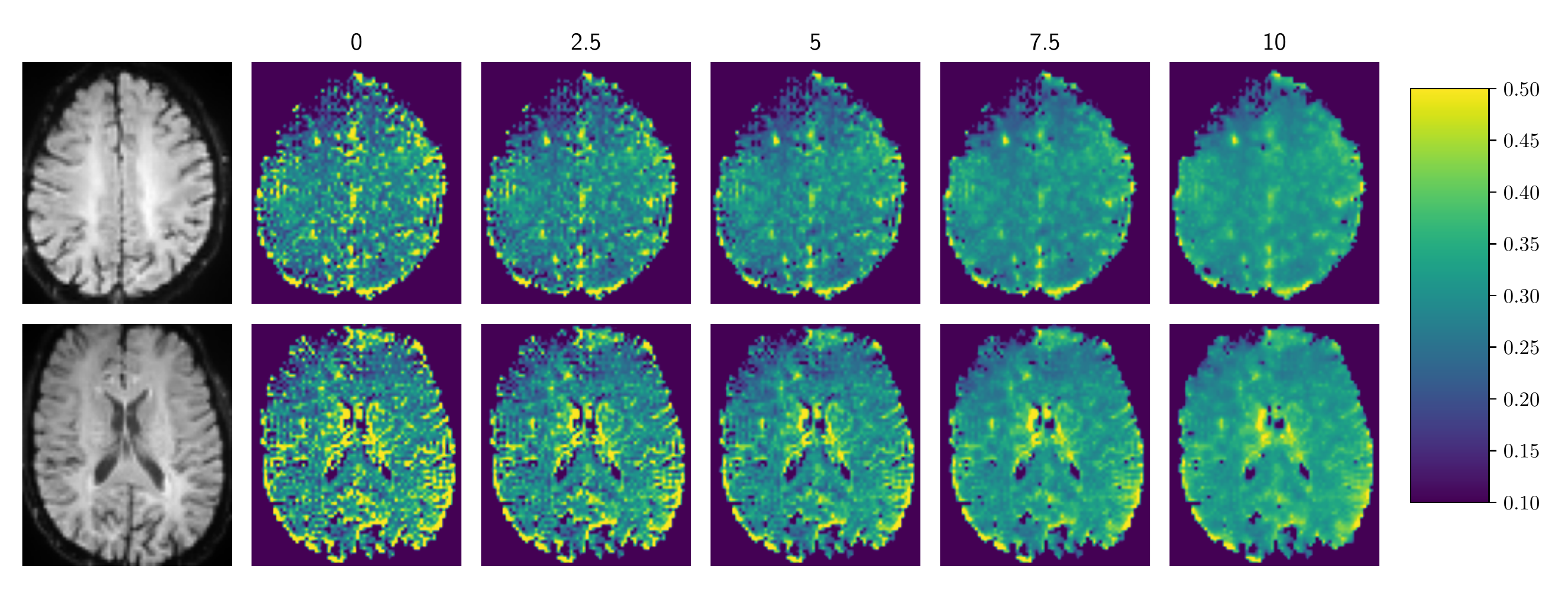}
    \caption{Demonstration of how the inferred $\oef$ maps change as the smoothness regularisation weight increases. \label{fig:smoothness}}
\end{figure}
We begin our analysis by investigating the effect of modifying the smoothness weight $\lambda$  on the resulting predictions. Results are illustrated in Figure \ref{fig:smoothness}.  We find that increasing $\lambda$ leads to smoother and more interpretable OEF maps, with a small decrease in ELBO. This is expected as the ELBO, in this case, is a voxelwise measure describing how well the image data is explained and how closely the prior is accorded with. Given the ambiguities of inferring OEF from this data, spatial smoothness introduces a useful degree of regularity in our inference. The differences in $\dbv$ and $\rp$ maps is very small, hence these are not shown. 

We select $\lambda = 5.0$ for all future experiments as it provides a good compromise between smooth parameter maps and preserving features.

\subsection{Effect of prior distribution}\label{sec:prior_dist}
\begin{figure}[!hbt]
    \begin{subfigure}[b]{0.45\textwidth}
    \centering
    \includegraphics[width=\columnwidth]{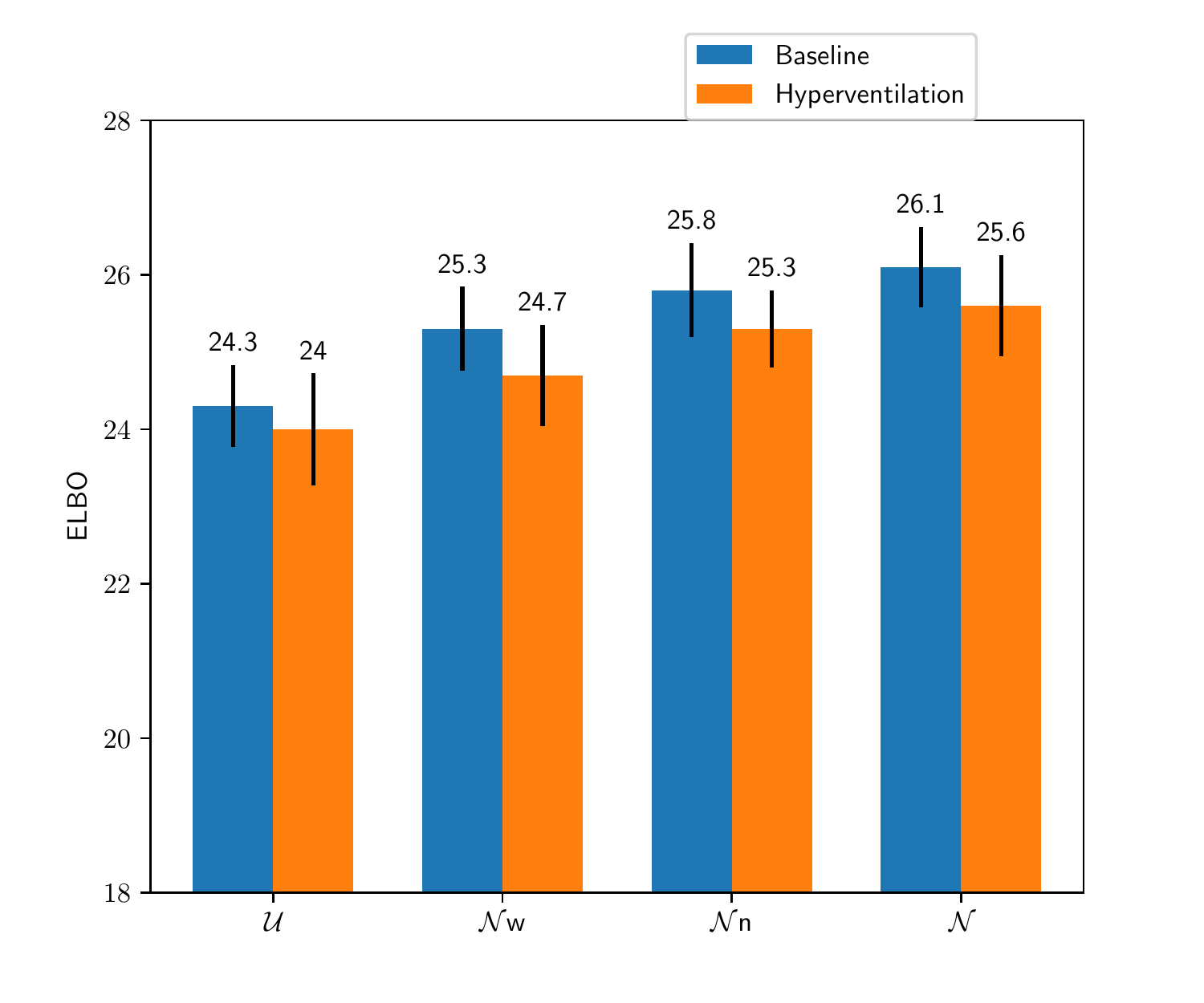}
    \caption{}
    \end{subfigure}
    \begin{subfigure}[b]{0.45\textwidth}
    \centering
    \includegraphics[width=\columnwidth]{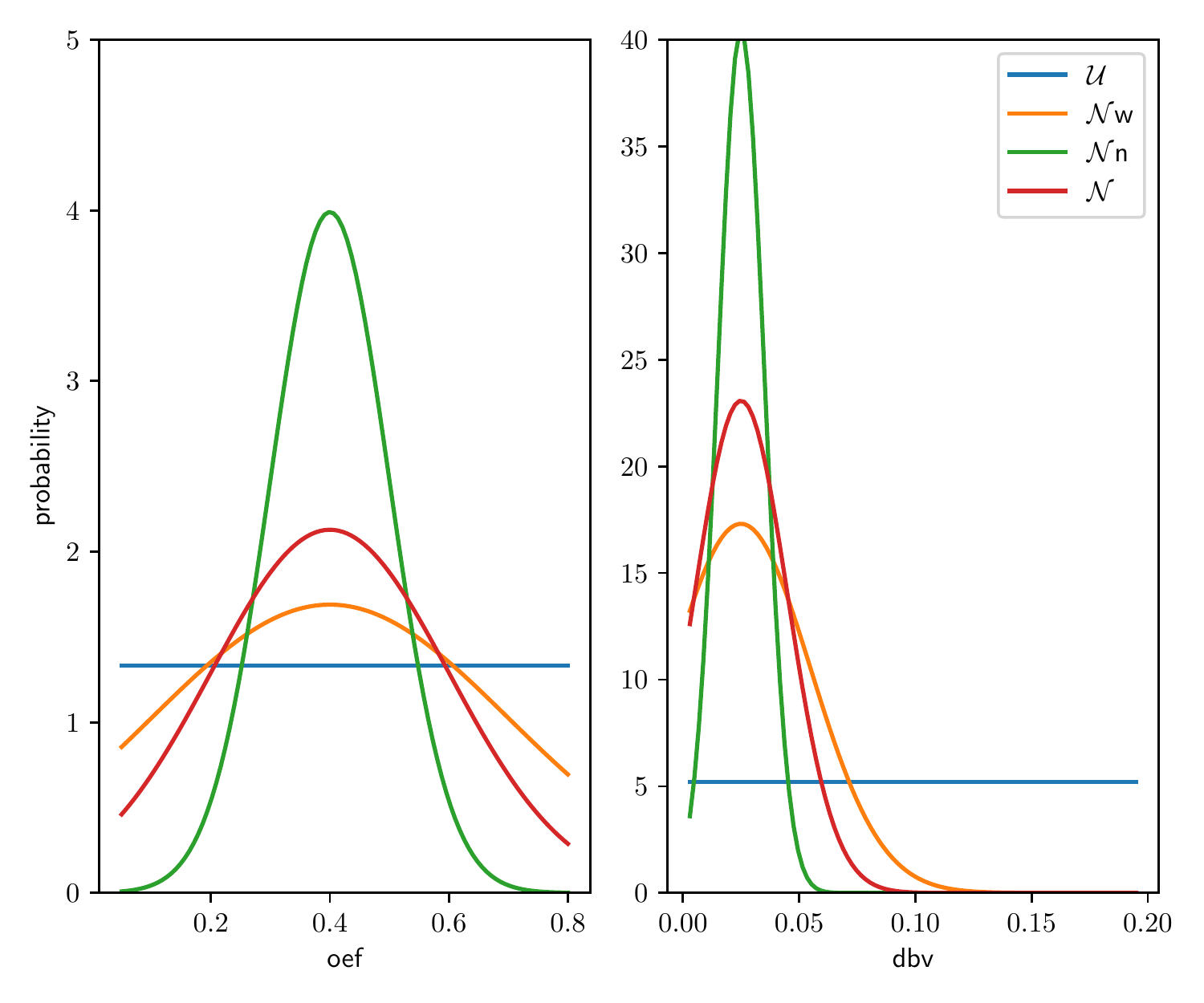}
    \caption{}
    \end{subfigure}
    \begin{subfigure}[b]{1.0\textwidth}
    \centering
    \includegraphics[width=\columnwidth]{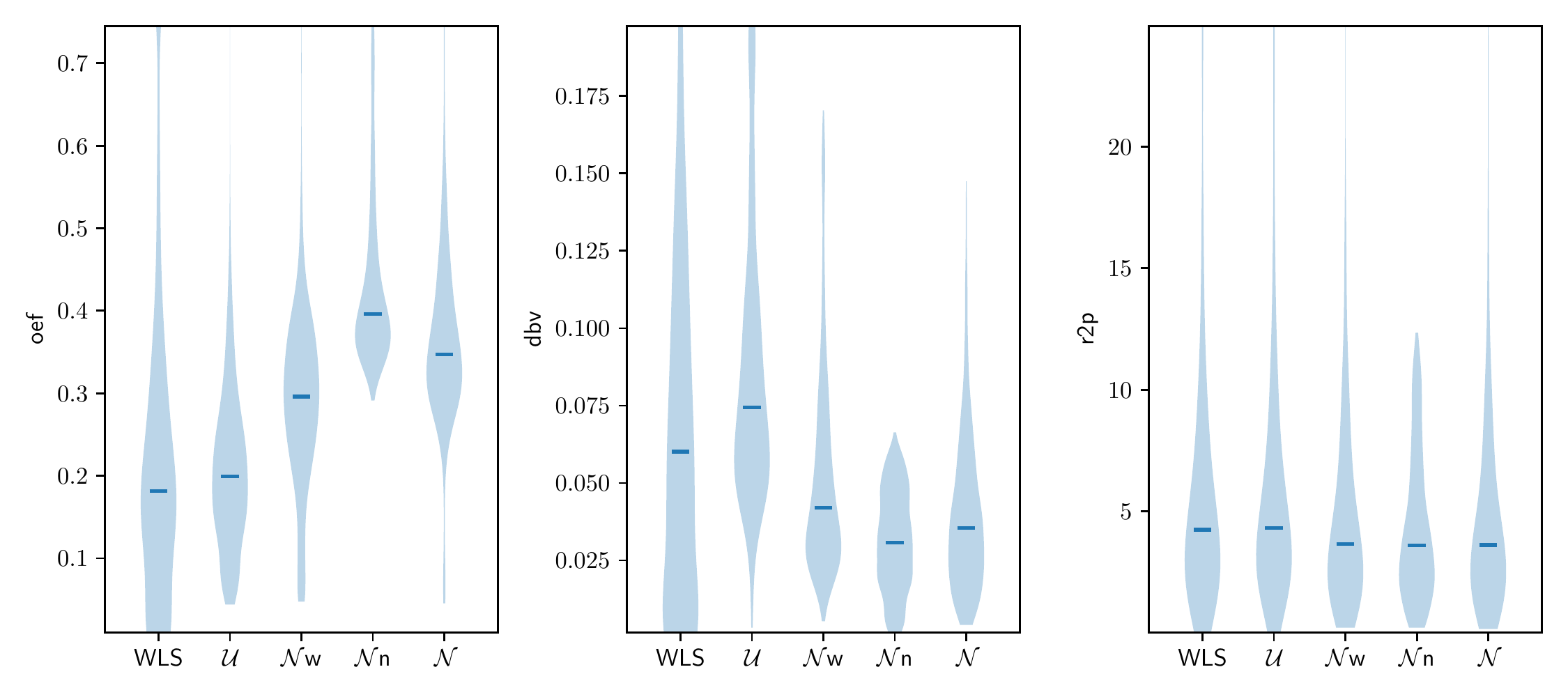} 
    \caption{ \label{fig:prior_oef_dbv}}
    \end{subfigure}
    \caption{The mean ELBO and std-deviation across subject-wise means for baseline, and hyperventilation conditions for grey matter voxels (a). Higher ELBO corresponds to better explanation of the original image data and closer accordance with the synthetically trained prior. b) Illustrates the OEF and DBV distributions used. 
    c) Shows the $\oef$ and $\dbv$ distribution at baseline condition in the gray matter for different distributions of synthetic data. WLS refers to weighted least squares inference, as given in \citep{stone2017streamlined}.
    $\mathcal{N}$ corresponds to synthetic data distributions from $\oef \sim \truncnorm(40\%, 20\%^2), \dbv \sim \truncnorm{N}(2.5\%, 2\%^2)$, $\mathcal{N}$w corresponds to $\oef \sim \truncnorm(40\%, 30\%^2), \dbv \sim \mathcal{N}(2.5\%, 3\%^2)$, $\mathcal{N}$n corresponds to $\oef \sim \truncnorm(40\%, 10\%^2), \dbv \sim \truncnorm(2.5\%, 1\%^2)$, and $\mathcal{U}$ corresponds to uniform distributions $\oef \sim \mathcal{U}(5\%, 80\%), \dbv \sim \mathcal{U}(0.3\%, 25\%)$. \label{fig:prior_elbo}}
\end{figure}
As can be seen in Figure \ref{fig:prior_elbo}, the choice of synthetic sample distribution has a substantial impact on the model's ability to predict the study image data as measured by the ELBO. Our choice of prior means for $\oef=40\%$ and $\dbv=2.5\%$ was supported from previous literature that found an \oef\ of $41 \pm 9 \%$ using Oxygen PET~\citep{derdeyn2001comparison} and $38.3 \pm 5.3$ using qBOLD \citep{He2007}. \dbv\ was estimated at $1.75 \pm 0.13 \%$ using qBOLD \citep{He2007}, $2.18 \pm 0.41$ in \citep{blockley2013analysis}, and was estimated at $3.1 \pm 0.5 \%$ using interleaved qBOLD \citep{lee2018interleaved}. We experimented with the standard deviations for these and found that $\oef \sim \truncnorm(40\%, 20\%^2)$ and $\dbv \sim \truncnorm(2.5\%, 2\%^2)$, provides a good compromise in terms of ELBO and tightness of the distribution, where $\truncnorm$ refers to a truncated Normal distribution to avoid values outside $[0\%,100\%]$. Interestingly, we note that in all cases the hyperventilation study data was less well predicted than the baseline dataset. We consider this in more detail in section \ref{sec:uncertainty}.

In Figure \ref{fig:prior_oef_dbv} we observe substantial differences in the distribution of $\oef$ and $\dbv$ depending on the choice of synthetic sampling distribution.  However, the distribution of $\rp$ does not alter significantly depending on these choices. 

These experiments serve to illustrate the intrinsic ambiguity in these data, i.e. there are a distribution of possible combinations of OEF and DBV that can explain any voxel reasonably well. Using a biologically motivated synthetic sampling distribution allows us to express our prior physiological knowledge, and prefer plausible values for OEF and DBV.



\subsection{Effect of modelling covariance between OEF and DBV}
As described in section \ref{sec:param_dist}, the approximate distribution of $\oef$ and $\dbv$ are modelled by scaled logit-Normal distributions. We experimented with either predicting diagonal or full covariance matrices, $\sigmafft$. Full covariance matrices describe the joint distribution of $\oef$ and $\dbv$ at each voxel, whereas the diagonal matrix assumes independence. Figure \ref{fig:diag_elbo} in the appendix shows that the full covariance matrix leads to improved predictions of the data as given by the ELBO, where we find a gray matter voxelwise mean of $26.1 \pm 0. 52$ for correlated \oef\ and \dbv\ and $24.2 \pm 0.64$ for independent distributions. We also note that the inferred mean \oef\ and \dbv\ values are similar for both methods..

\subsection{Predictive uncertainty and identifying outliers}\label{sec:uncertainty}
\begin{figure}[!hbt]
    \begin{subfigure}[b]{0.49\textwidth}
    \centering
    \includegraphics[width=\columnwidth]{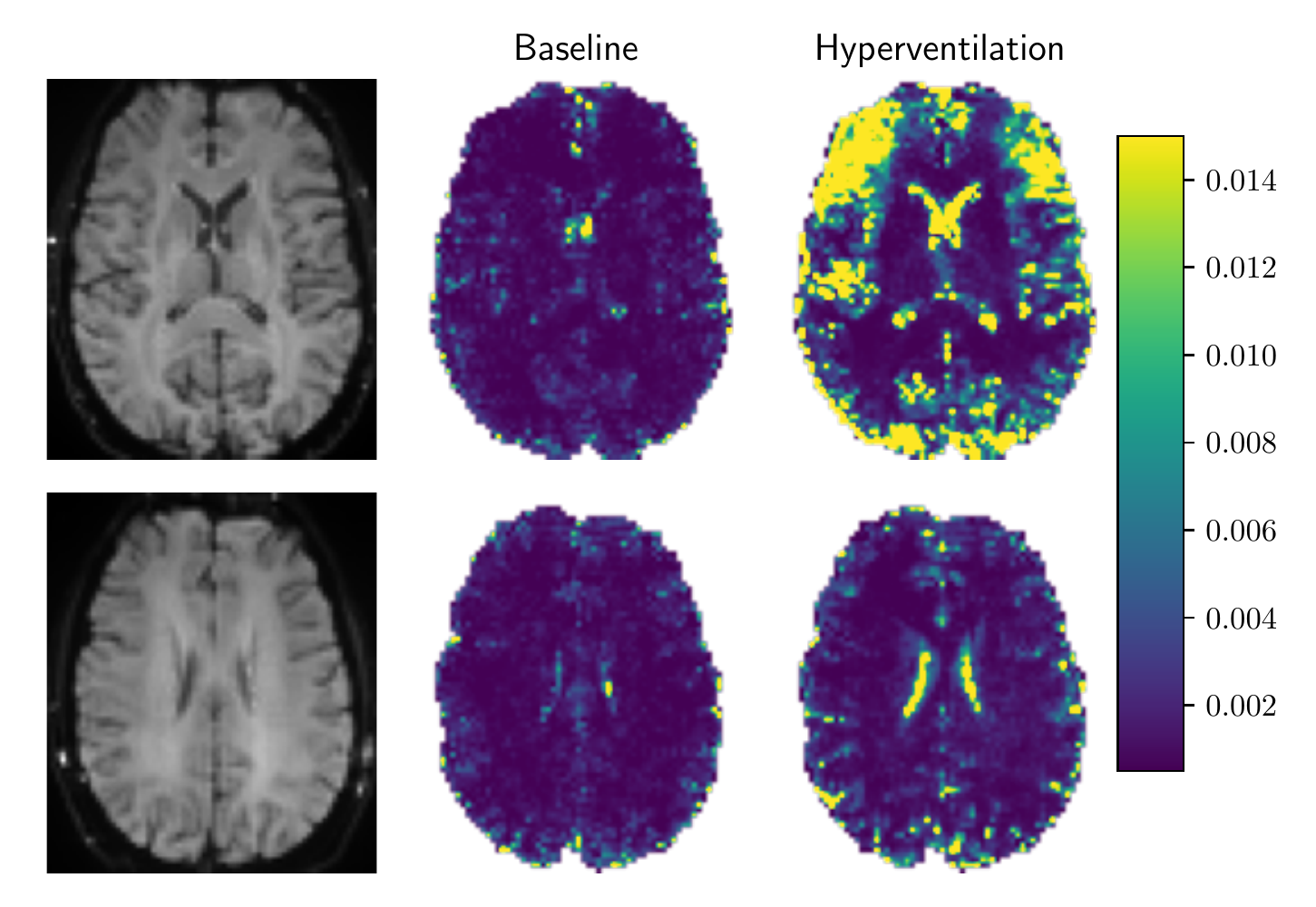}
    \caption{}
    \end{subfigure}
    \begin{subfigure}[b]{0.49\textwidth}
    \centering
    \includegraphics[width=\columnwidth]{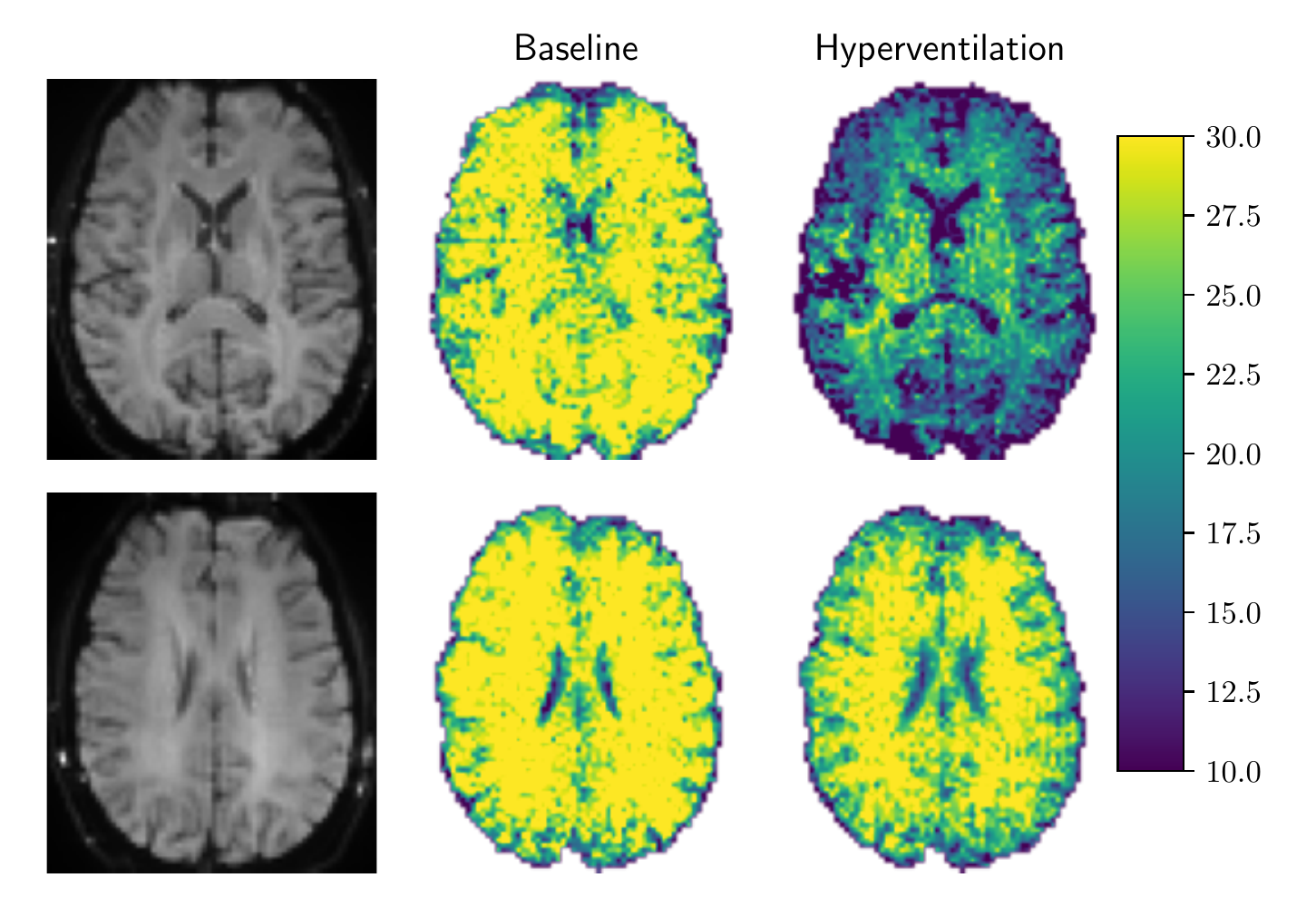}
    \caption{}
    \end{subfigure}
    \caption{Illustration of the predicted uncertainty of OEF (std-dev given in a) and the ELBO (given in b). The top row corresponds to the subject with artefacts in the hyperventilation scan, the bottom row is another subject. We can clearly see substantially higher uncertainty and lower ELBO in the scan with artefacts. \label{fig:uncertainty}}
\end{figure}
We have two means of identifying problematic data, the predictive uncertainty and the ELBO values. Figure \ref{fig:uncertainty} illustrates maps of predictive uncertainty and ELBO on two subjects from the study dataset. For the subject with artefacts in their hyperventilation qBOLD scan, we can easily identify the issues using either of these measures. We also observe lower ELBO in hyperventilation scans (GM mean $25.6 \pm 0.65$ compared to $26.1 \pm 0. 52$) at baseline; this is particularly visible around the cortex and near regions of CSF, see Figure \ref{fig:uncertainty} b) bottom right for an example. We hypothesise that this is due to additional subject motion and changing partial volumes over the acquisition during hyperventilation.

\subsection{Choice of forward model}
We consider using a one, or two compartment model (including the contribution from venous blood described in Section 2.3) with either the asymptotic or full signal equations.  We find that using the full signal model has a substantial impact in explaining the data, with an increase in gray mater voxelwise EBLO from $23.0 \pm 0.65$ to $25.7 \pm 0.56$ with one compartment and $22.5 \pm 0.60$ to $26.1 \pm 0.52$ with two compartments. The difference in ELBO between 1 and 2 compartment models is reasonably small, and the parameter distributions are very similar. See Figure \ref{fig:forward_elbo} in the appendix for further details.

\subsection{Comparing Inference Strategies}
We compare four inference variants: weighted least squares (WLS) \citep{stone2017streamlined}, our model trained only on synthetic data (Synth), and variational inference without and with spatial regularisation, VI and VI + TV respectively. The distribution of gray matter values for $\oef$, $\dbv$ and $\rp$ for the 4 methods in Table \ref{tab:param_table}.

\begin{table*}[]
    \centering
    \begin{tabular}{c|c|c|c|c|c|c}
        \hline \hline
         Method & \oef\ & \oef-hyp & \dbv\ & \dbv-hyp & $\rp$ & $\rp$-hyp  \\ \hline\hline
         WLS &  $23.2 \pm 20.9$  &  $25.5 \pm 21.7$ &  $7.65 \pm 7.07$  &  $8.03 \pm 6.88$ &  $6.2 \pm 6.3$  &  $6.8 \pm 5.3$ \\
         \hline
         Synth & $38.2 \pm 13.4$  &  $41.9 \pm 13.4$ & $4.45 \pm 2.55$  &  $4.80 \pm 2.19$ &$5.5 \pm 5.1$  &  $6.2 \pm 4.4$ \\ \hline
         VI & $35.5 \pm 15.8$  &  $39.6 \pm 15.1$ & $4.37 \pm 2.78$  &  $4.77 \pm 2.35$ & $5.3 \pm 5.5$  & $ 6.0 \pm 4.7$ \\ \hline
         VI + TV &  $35.1 \pm 13.0$  &  $39.1 \pm 12.1$ & $4.21 \pm 2.60$  &  $4.60 \pm 2.17$ &  $5.2 \pm 5.4$  &  $5.9 \pm 4.6$ \\ \hline
         \hline
    \end{tabular}
    \caption{Gray-matter parameter means and standard deviations for different methods at baseline and during hyperventilation.}
    \label{tab:param_table}
\end{table*}

Our learned approaches, given the choice of prior distributions, produce values for OEF and DBV that are in a more plausible range than WLS. The Synth results show closer adherence to the prior distribution in terms of \oef\, as this model is heavily regularised. VI leads to higher standard deviations, whereas the use of total variation regularisation (VI+TV) narrows the inferred \oef\ and \dbv\ distributions considerably. $\rp$ is reasonably stable across learned inference methods. See Figure \ref{fig:baseline_distribution} in the appendix for violin pots of the distributions at baseline.

\begin{figure*}[!hbt]
    \centering
    \begin{subfigure}[b]{0.7\textwidth}
    \centering
    \includegraphics[width=\columnwidth]{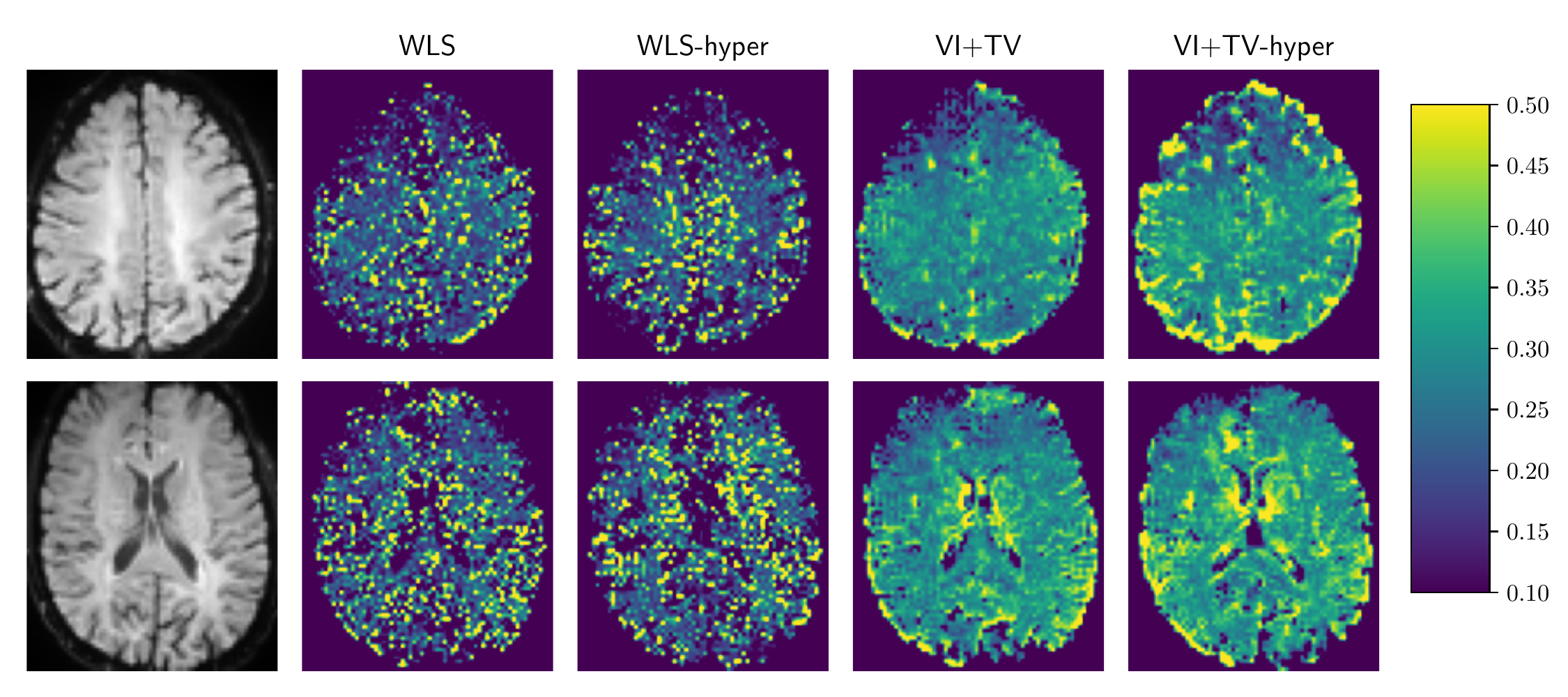}
    \caption{}
    \end{subfigure}
    \begin{subfigure}[b]{0.7\textwidth}
    \centering
    \includegraphics[width=\columnwidth]{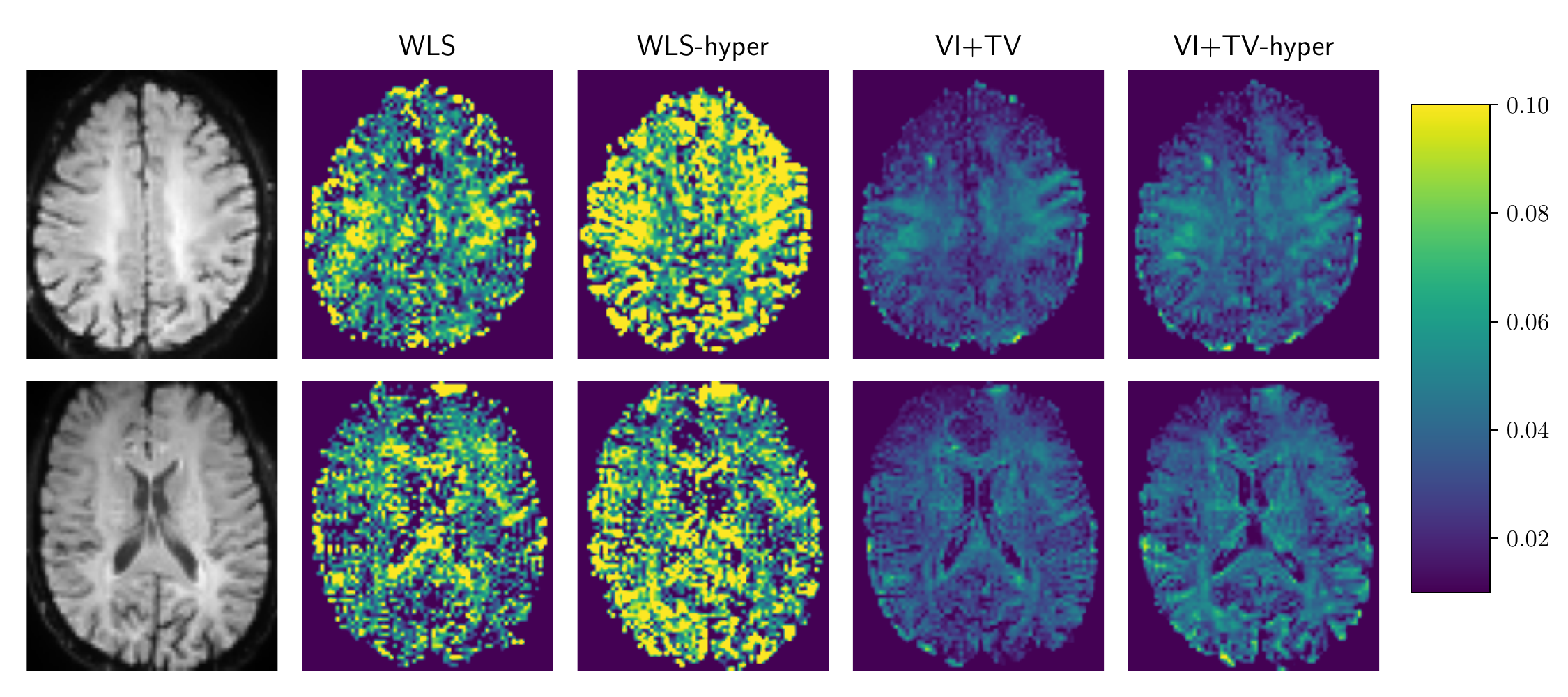}
    \caption{}
    \end{subfigure}
    \begin{subfigure}[b]{0.7\textwidth}
    \centering
    \includegraphics[width=\columnwidth]{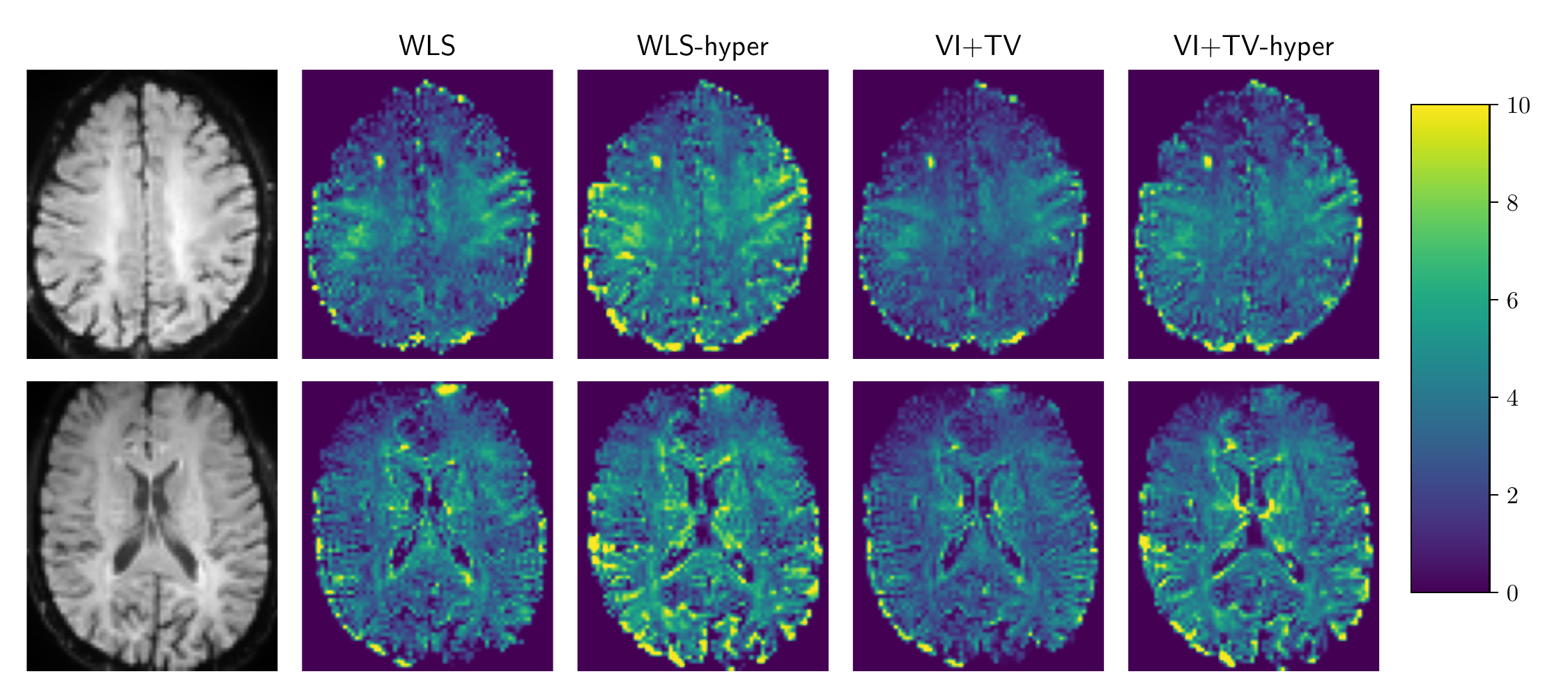}
    \caption{}
    \end{subfigure}
    \caption{Example \oef\ maps (a), \dbv\ maps (b) and $\rp$ maps (c) for baseline and hyperventilation using WLS or our proposed method. We observe that VI+TV leads to substantially more interpretable maps than WLS, where the differences between baseline and hyperventilation are clearly visible.}
    \label{fig:example_maps}
\end{figure*}
Example parameters maps are provided in Figure \ref{fig:example_maps} for WLS and VI+TV. WLS maps are particularly noisy for $\oef$ and $\dbv$, whereas VI+TV leads to smoother maps.

\subsection{Hyperventilation Study}
\begin{figure*}[!hbt]
    \centering
    \begin{subfigure}[b]{0.8\textwidth}
    \centering
    \includegraphics[width=\columnwidth]{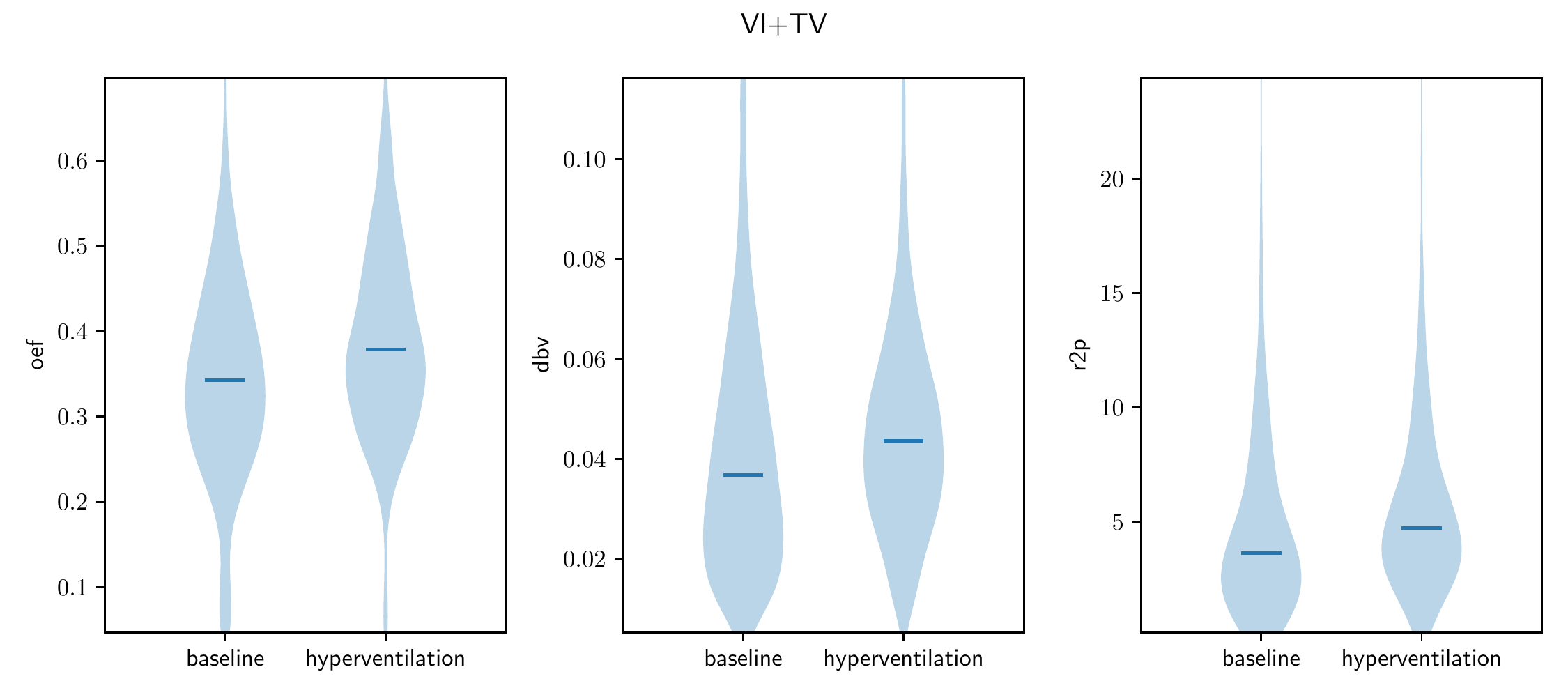}
    \caption{}
    \label{fig:vitv_condition_dist}
    \end{subfigure}
    \begin{subfigure}[b]{0.8\textwidth}
    \centering
    \includegraphics[width=\columnwidth]{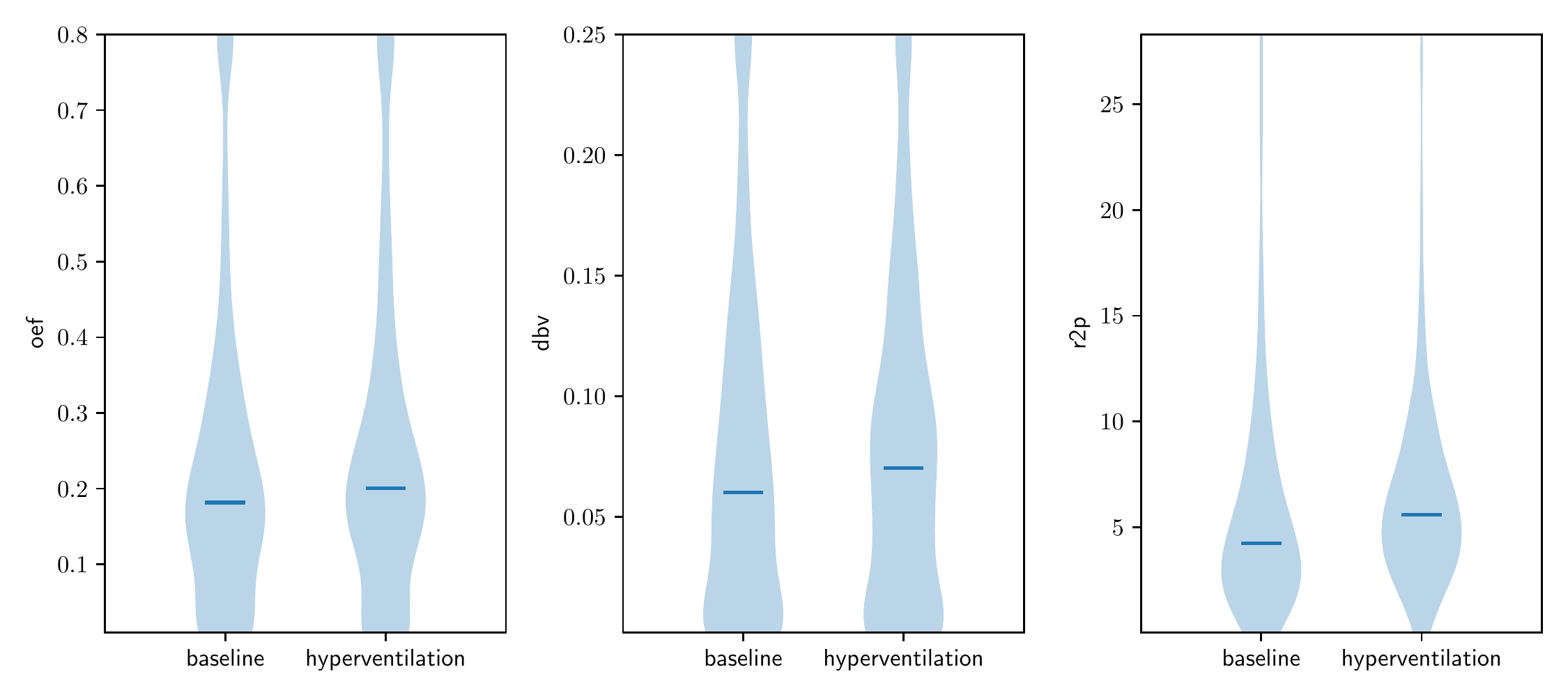}
    \caption{}
    \label{fig:wls_condition_dist}
    \end{subfigure}
    \caption{Grey matter $\oef$, $\dbv$ and $\rp$ parameter distributions for baseline and hyperventilation conditions for either our VI+TV approach (a), or WLS (b). VI+TV demonstrates distinct differences between the two populations for all parameters. WLS infers heavy-tailed distributions for $\oef$ and $\dbv$.}
\end{figure*}

Finally, we compare the baseline and hyperventilation conditions in our study dataset. Examining the individual maps in Figure \ref{fig:example_maps} and numbers in table \ref{tab:param_table}, there is noticeable increase in \oef\ \dbv\ and $\rp$ during hyperventilation. Figure \ref{fig:vitv_condition_dist} provides a more detailed inspection of the parameter distributions for VI+TV, we can see for our optimal approach that there are distinct differences in the grey matter between the two populations for all the parameters. In comparison in Figure \ref{fig:wls_condition_dist}, WLS exhibits much heavier tailed distributions for $\oef$ and $\dbv$ with less obvious differences.


\subsection{Voxelwise Comparisons}
\begin{figure}[!hbt]

    \centering
    \begin{subfigure}[b]{0.48\textwidth}
    \centering
    \includegraphics[width=0.6\columnwidth]{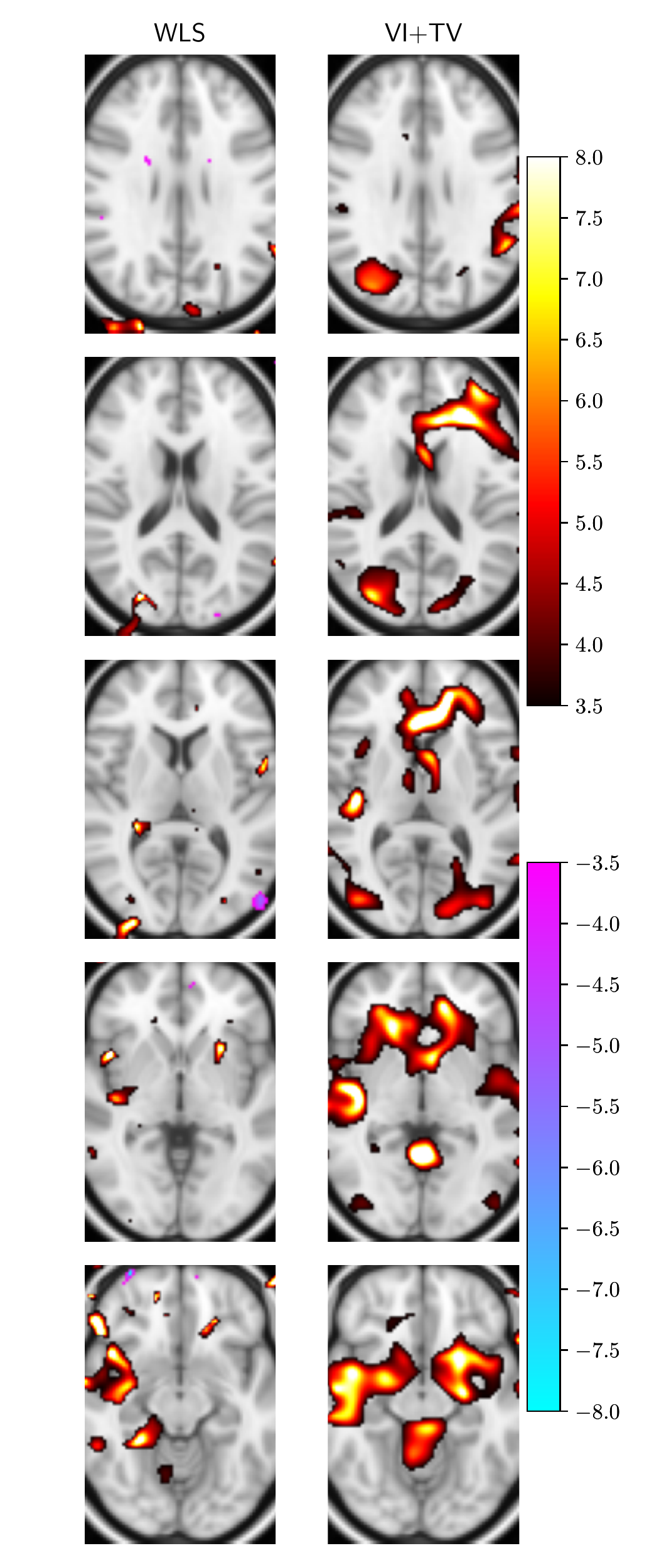}
    \caption{}
    \label{fig:t_stat_oef}
    \end{subfigure}
    \begin{subfigure}[b]{0.48\textwidth}
    \centering
    \includegraphics[width=0.6\columnwidth]{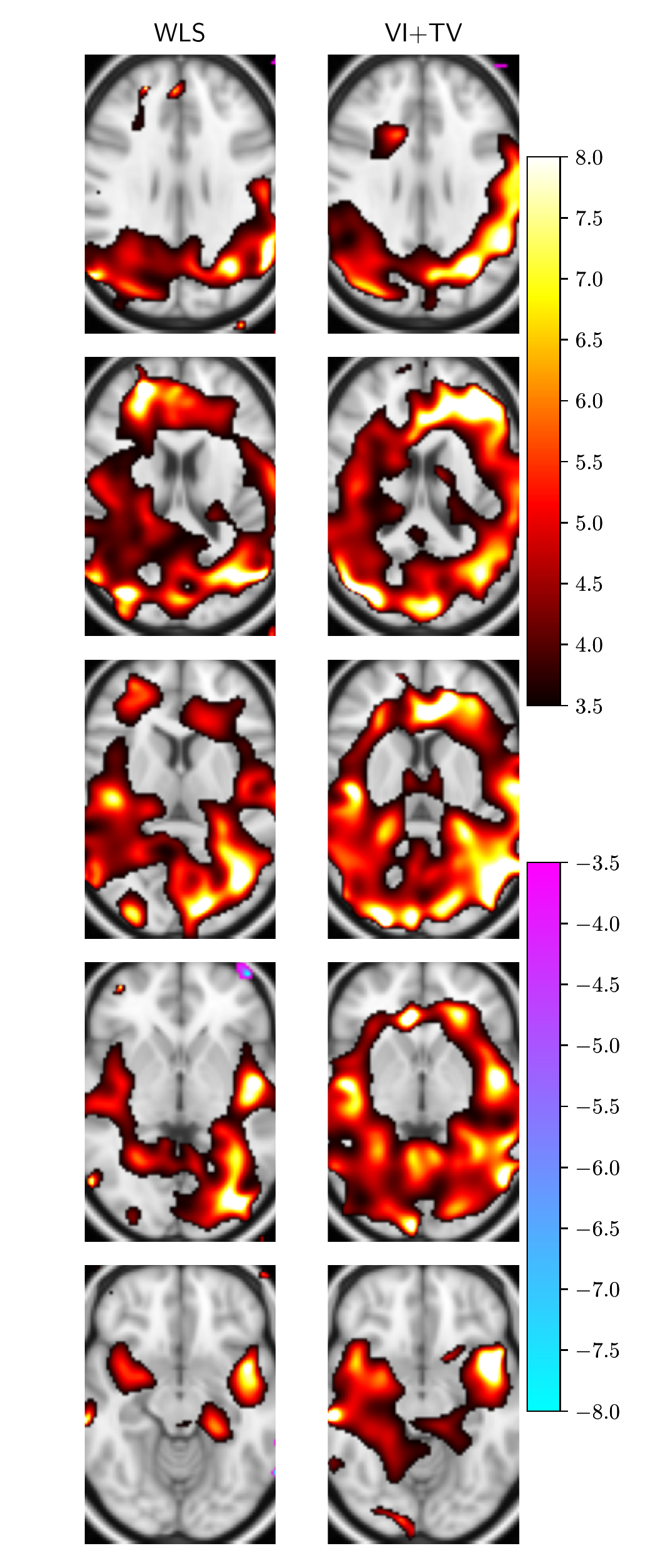}
    \caption{}
    \label{fig:t_stat_r2p}
    \end{subfigure}
    \caption{Uncorrected t-statistic maps of \oef\ (a) and $\rp$ (b) calculated by a paired t-test.  The $\rp$ maps are similar for both methods, but only our approach identifies substantial clusters of voxels with increased \oef.  These maps illustrate substantial voxelwise increases in \oef\ and $\rp$ across the brain during hyperventilation.  \dbv\ showed no voxelwise significant differences despite the overall differences across the gray matter, as shown in Figure \ref{fig:vitv_condition_dist}.}
\end{figure}
Data was spatially normalised to the MNI152 template using the ``fsl\_anat" tool in FSL. This performed linear and non-linear registration of our study subjects to a normalised atlas. To align the inferred parameter maps a combination of linear registration to the T1, and non-linear EPI distortion correction using the FieldMap toolkit in SPM~\citep{andersson2001modeling}, where the field maps were derived from the qBOLD phase data acquired at different echo times.
The spatially normalised maps were smoothed with a Gaussian kernel with a FWHM of 6mm. Two-way paired t-test were run between the two conditions. Figures \ref{fig:t_stat_oef} and \ref{fig:t_stat_r2p} present uncorrected t-statistc maps for visual comparison. We find that our proposed method provide a substantial improvement in localising differences in $\oef$ compared to WLS.
We note that DBV does not identify any significant voxelwise changes in hyper-ventilation, despite the differences given across the population in Figure \ref{fig:vitv_condition_dist}.


\subsection{Adaptability to new data}

To demonstrate the ability of our model to be applied to alternative data with different parameters, we experimented with the 7 acquisitions collected in \citep{stone2017streamlined}\footnote{Data available from: https://ora.ox.ac.uk/objects/uuid:177afade-8599-4d9c-959e-26e1426ec486}. These data were acquired with a coarser resolution, 3.75$mm$ and a denser sampling in $\tauvec$ (every $4ms$ from -28 to 64).  Despite the small set of samples, we found that the only change required for successful training was to increase the weight decay regularisation during the prior training phase ($5e^{-4}$). We hypothesise this was required to avoid overfitting to the more detailed input data as there are 24 images rather than 11. We present predicted \oef\ and \dbv\ maps in Figure \ref{fig:alt_data} in the appendix, again noting that the derived $\rp$ maps appear very similar to those inferred using WLS.

\section{Discussion}
This work has introduced adaptive and informative voxelwise prior information using regularised models trained on synthetic data.  We found that the choice of OEF/DBV distribution chosen for generating the synthetic signals has a strong impact on the final inferred distribution. Accordingly, the choice of sampling distribution can be considered the population prior distribution for OEF and DBV. In this paper, the distributions were chosen based on literature review. However, the full influence on inferred parameters in the case of disease requires further investigation. We would expect that the implied bias of the prior may dampen any inferred deviations. As the presented framework is fast to train, and requires minimal real data, external knowledge of the expected parameter distribution can be easily incorporated.

Although we do not present results here, we did also experiment with simpler global priors on \oef\ and \dbv\ that follow a scaled logit-Normal, or mixture of scaled logit-Normal distributions that were jointly inferred based on the training dataset. However, all of these experiments led to poor model fitting. This was probably related to our data masking procedure, where only voxels that were definitely outside of the brain were excluded, and perhaps such a global approach might be more effective with tissue specific priors. 

Streamlined qBOLD quantification of \oef\ and \dbv\ may be biased by differences due to susceptibility to the unknown iron content in neuroinflamation, or in the deep gray matter of the basal ganglia, where we observe strong signals.  Nonetheless, even under such conditions streamlined qBOLD may provide useful assessment of within-region OEF differences, temporal changes, or between experimental conditions as demonstrated in this study. Reassuringly, the observed OEF signal increase during hyperventilation is perfectly in line with predictions~\citep{diringer2000no}\citep{hutchinson2002correlation} 

Based on the close coupling between CBF and CMRO2, and their apparently  similar regional distributions,  OEF has been hypothesized to be uniform throughout the brain \citep{coles2006intersubject}\citep{raichle2001default}. Recently, however, more precise observations~\citep{henriksen2021regional} revealed a variable pattern of distribution of OEF between brain areas, reflecting regional specific alterations in the relationships between delivery and consumption of brain energy substrates. This regional variability in OEF values, despite accounting for only a small proportion of the total OEF variability (which for the most part is driven by  between-subject variations), is of great clinical relevance. Factors such as alterations in neurovascular uncoupling, use of alternate energy substrates and rate of aerobic glycolysis, could all contribute to alter the CMRO2/CBF relationship. The notion that some of these contributions might be regionally variable across the brain \citep{blazey2019quantitative}, makes the non-invasive study of the precise distribution of OEF in the living human brain particularly compelling. This approach could potentially unveil pathophysiological mechanisms underlying the unique vulnerability of specific brain areas to energy dysregulation and pathology, with great relevance for the definition of novel treatment targets for neurological and neuropsychiatric conditions, ranging from Stroke and Multiple Sclerosis, to Bipolar Disorder and Schizophrenia.

Other sources of bias include the assumed constant value for $\hct = 0.34$. Although $\hct$ can be measured directly from blood, it is expected that the $\hct$ in small vessels may be lower than that of general circulation \citep{fahraeus1931viscosity}, which is why several prior works on qBOLD \citep{He2007,stone2017streamlined} have used $\hct=0.34$ rather than $0.4$ as expected for general circulation used in \citep{Cherukara2019, berman2018transverse}. 
As $\hct$ is closely correlated with haemoglobin concentration, an alternative option is to exploit the dependency of T1 on the concentration of haemoglobin to use MRI-based methods to estimate inter-individual variations \citep{xu2018accounting}.

\section{Conclusions}
This paper has introduced an efficient new amortized inference framework for inferring quantitative model parameters from MRI data, applied to qBOLD MRI. We have illustrated how synthetic datasets can be used to create informative prior distributions, which can then be incorporated into a variational Bayesian inference model. Our formulation enables any differentiable forward model for the data to be used, and we found that the full qBOLD signal equation explains the data better than the asymptotic approximation. To constrain \oef\ and \dbv\ to a realistic range, we described them jointly using a scaled multivariate logit-Normal distribution, which removes some of the impossible values inferred in previous works. The proposed method was shown to provide smoother and more realistic parameters maps for \oef\ and \dbv, which enables plausible voxelwise comparison of changes in hyperventilation.

Future work will consider integrating this approach with cerebral blood flow estimation from arterial spin labelling data \citep{alsop2015recommended}, which will permit straightforward inference of the cerebral metabolic rate of Oxygen~\citep{an2001quantitative}. 
Other directions of further research include more principled ways to integrate spatial smoothness priors, and experimenting with different compartment models to explain the data.

\textbf{Acknowledgments:} We acknowledge funding from the University of Brighton Rising Stars initiative and support from Brighton and Sussex Medical School. We thank Iris Asllani and Itamar Ronen for helpful discussions.

%% file: body/appendix.tex
\appendix
\newpage
\section{Architecture}
\begin{figure}[!h]
    \centering
    \includegraphics[width=0.7\columnwidth]{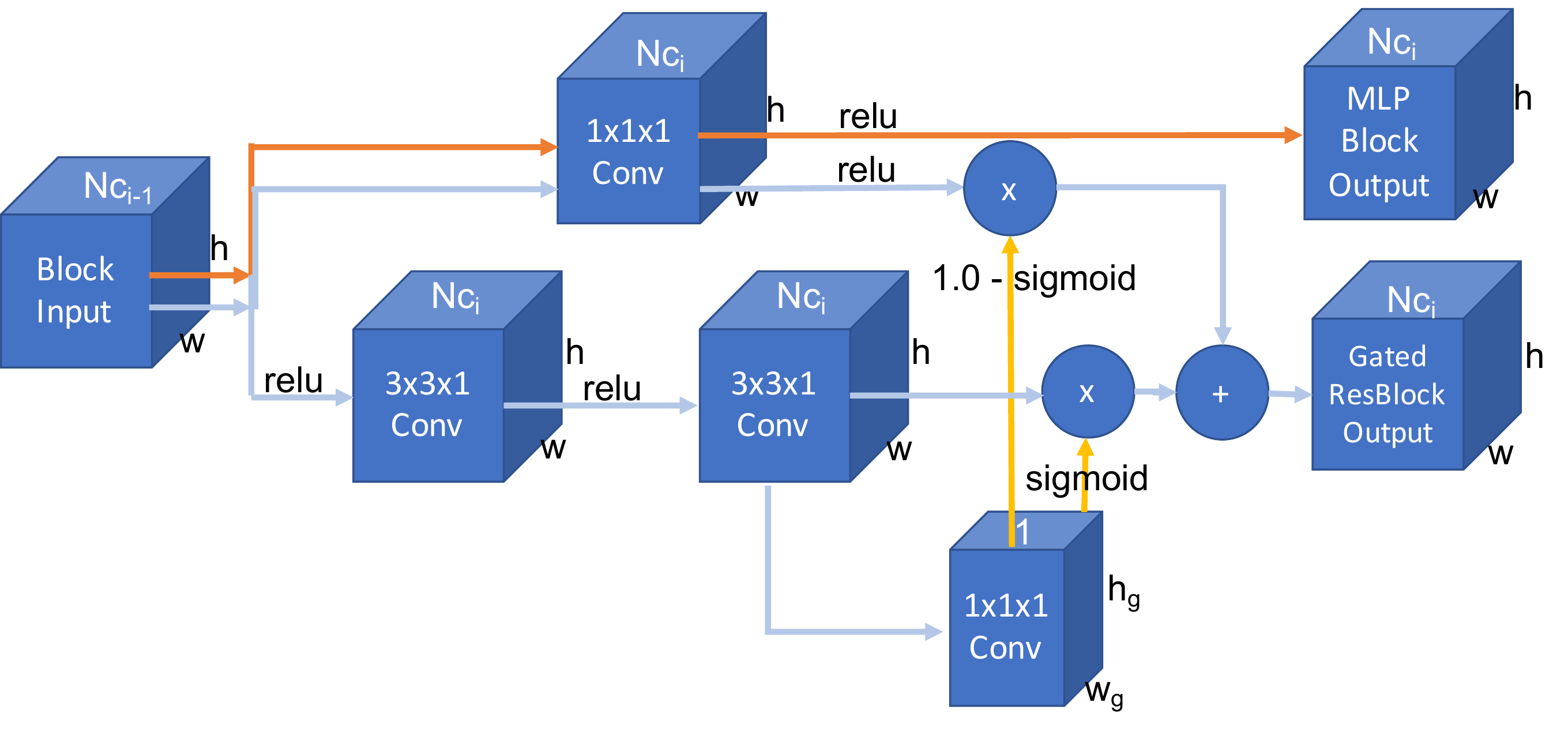}
    \caption{A graphical illustration of the convolutional blocks. The red line corresponds to the flow of information for the MLP block output, which is pre-trained using synthetic data. The light blue line corresponds to the gated residual convolutional block. The gating information is given by the yellow lines, and either produces a scalar (through global mean averaging) or a voxelwise weighting.}
    \label{fig:conv_block}
\end{figure}
\newpage
\section{Covariance modelling}
\begin{figure}[!h]
    \begin{subfigure}[b]{0.3\textwidth}
    \centering
    \includegraphics[width=\columnwidth]{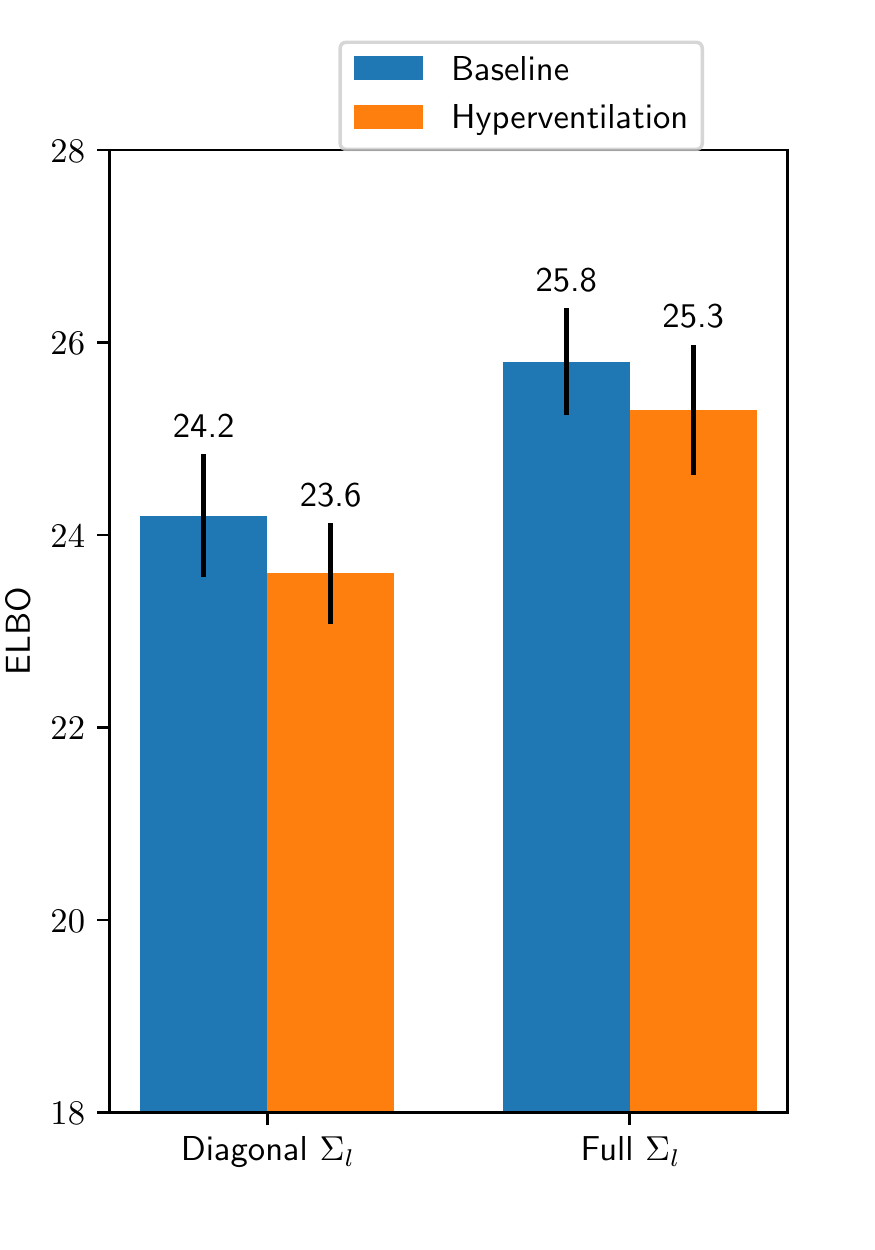}
    \caption{}
    \end{subfigure}
    \begin{subfigure}[b]{0.8\textwidth}
    \centering
    \includegraphics[width=\columnwidth]{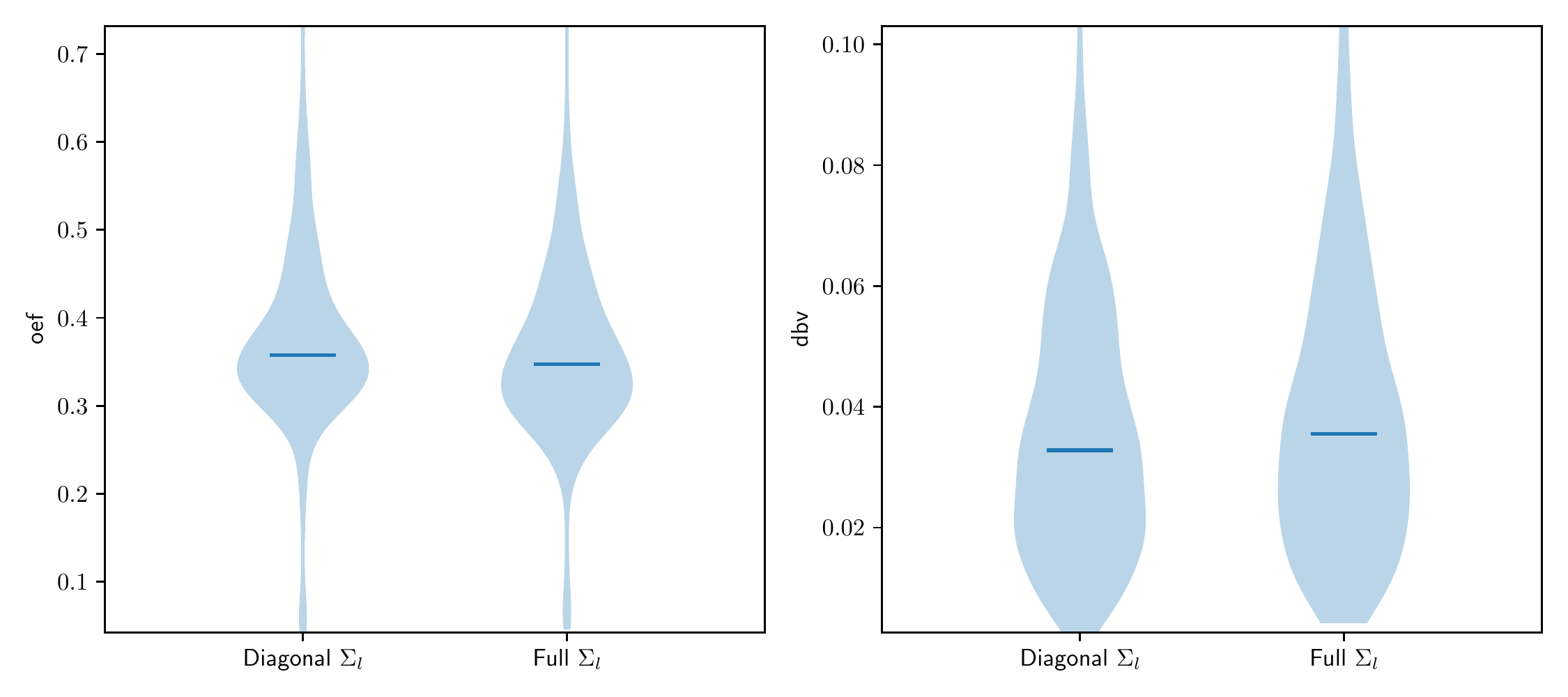}
    \caption{}
    \end{subfigure}
    \caption{ Results of using a diagonal or full covariance matrix, $\Sigma$ for $\oef$ and $\dbv$. a) Shows the mean and std-deviation across subject-wise ELBO means for baseline, and hyperventilation conditions in grey matter voxels. b) Show a violin plot of the grey matter distribution at at baseline condition in the gray matter. \label{fig:diag_elbo}}
\end{figure}
\newpage
\section{Choice of forward model}
\begin{figure}[!h]
    \centering
    \begin{subfigure}[b]{0.6\textwidth}
    \includegraphics[width=\columnwidth]{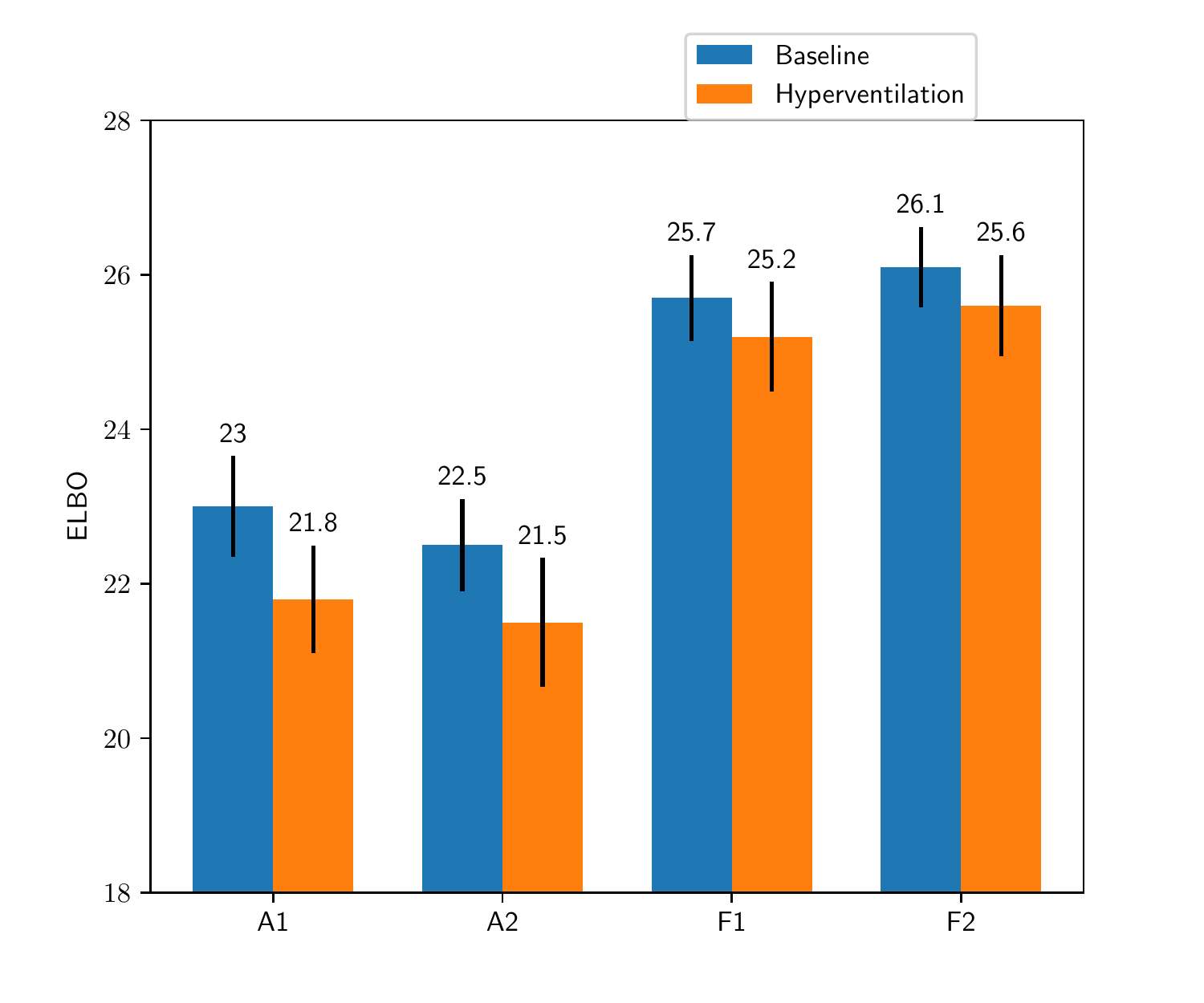}
    \caption{}
    \end{subfigure}
    \begin{subfigure}[b]{0.7\textwidth}
    \includegraphics[width=\columnwidth]{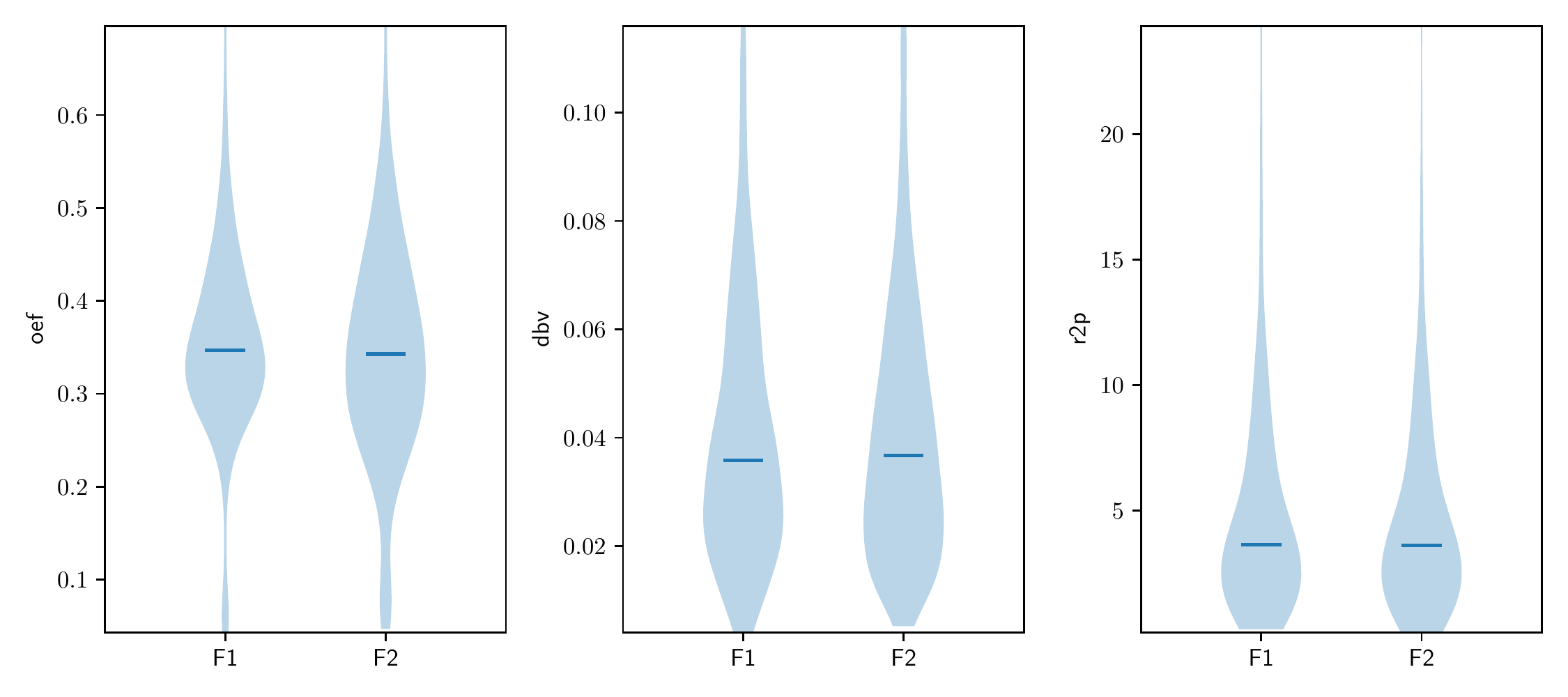}
    \caption{}
    \end{subfigure}
    \caption{a) ELBO, $\elbo$, for different forward models. A refers to asymptotic approximations, and F the full forward model. The number indicates the number of compartments. b) Difference in inferred parameter distribution at baseline for F1 and F2.}
    \label{fig:forward_elbo}
\end{figure}
\newpage
\section{Parameter distributions for different methods}
\begin{figure}[!hbt]
    \centering
    \includegraphics[width=\columnwidth]{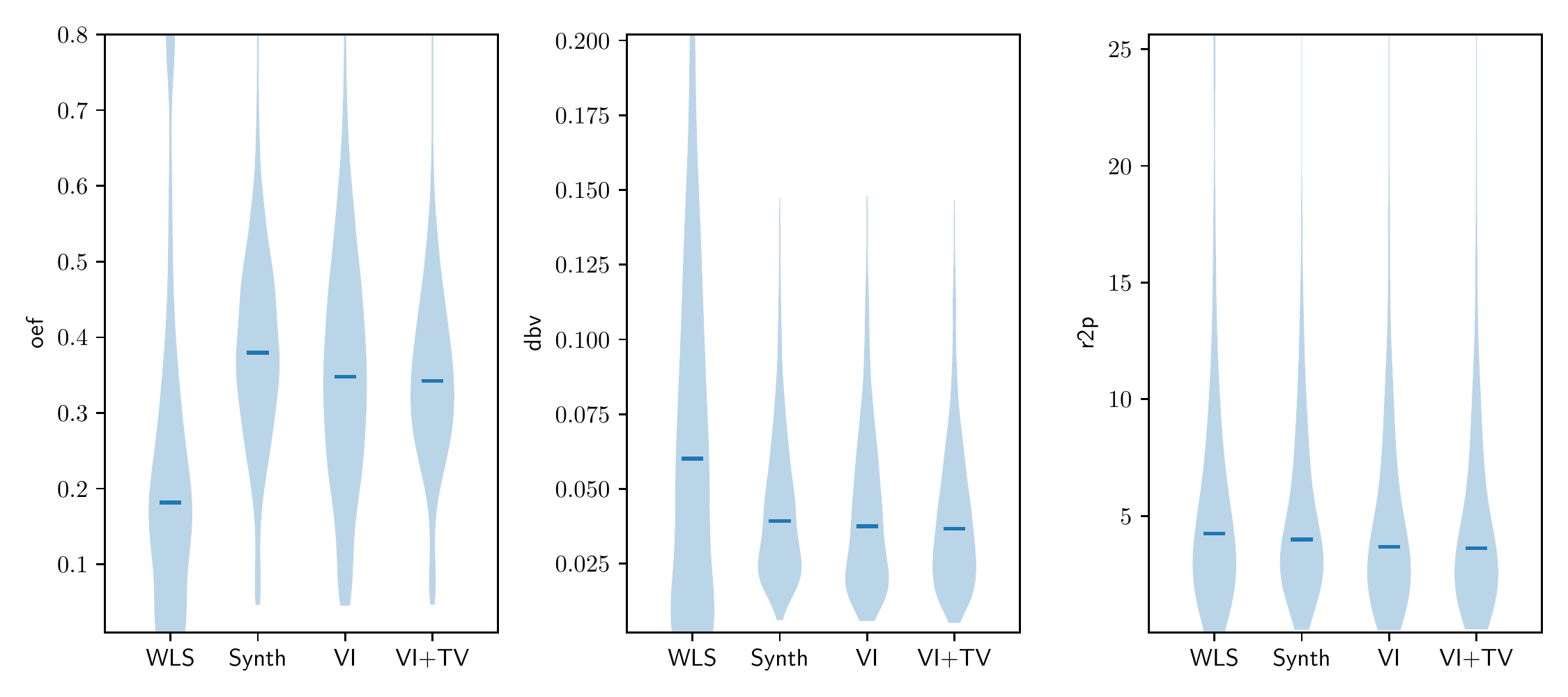}
    \caption{Distribution of baseline gray matter voxels for difference inference approaches. We see that the synthetically trained model estimates higher $\dbv$ and $\rp$ than the other methods, and WLS has heavy tails.}
    \label{fig:baseline_distribution}
\end{figure}
\newpage
\section{Adapatability to different data}
\begin{figure}[!h]
    \centering
    \begin{subfigure}[b]{0.45\textwidth}
    \centering
    \includegraphics[width=\columnwidth]{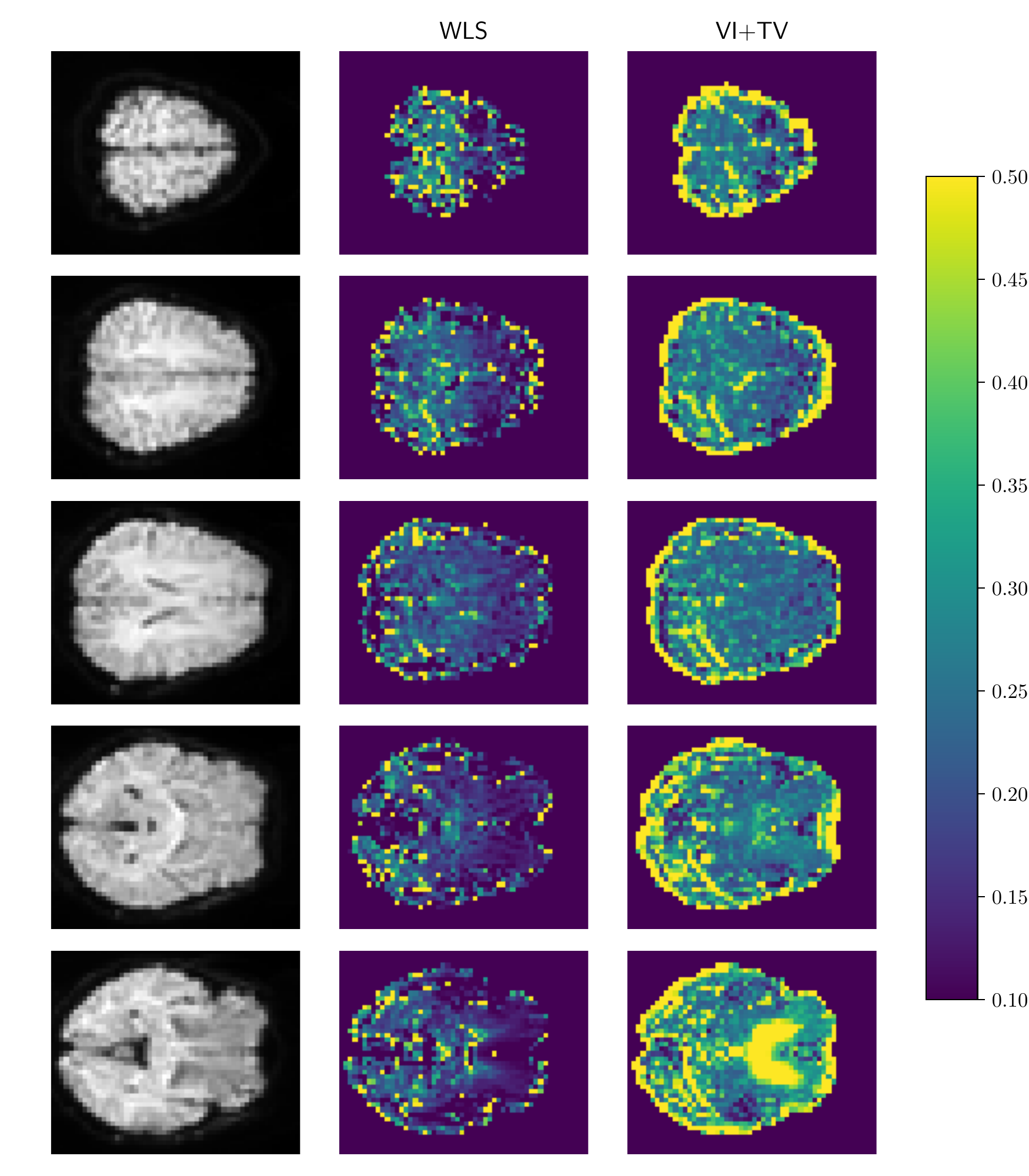}
    \caption{}
    \end{subfigure}
    \begin{subfigure}[b]{0.45\textwidth}
    \centering
    \includegraphics[width=\columnwidth]{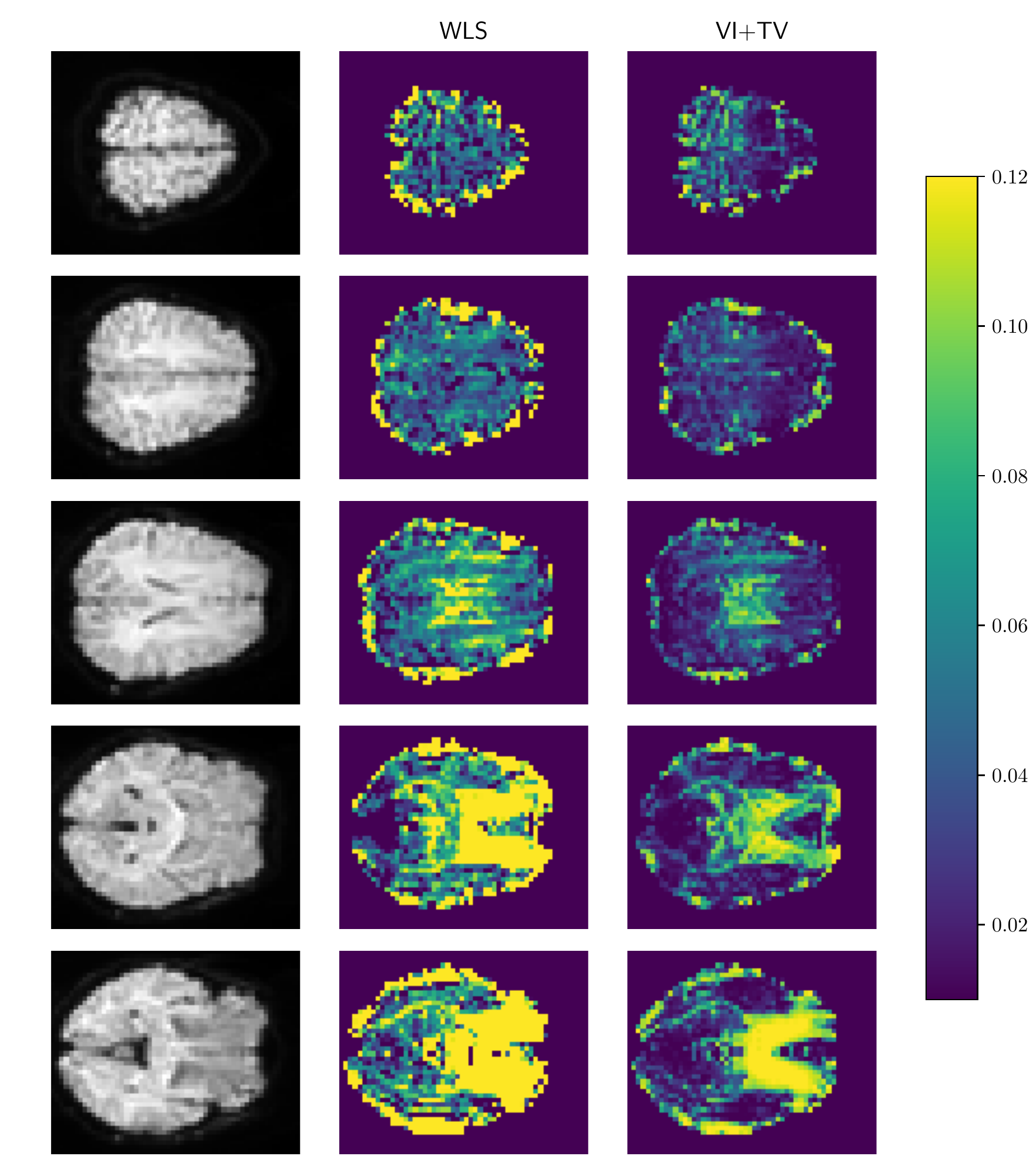}
    \caption{}
    \end{subfigure}
    \caption{Example $\oef$ maps (a) and $\dbv$ maps (b) for one held-out subject using the data from \citep{stone2017streamlined}. We observe similar patterns of $\dbv$, but with changes in scale. $\oef$ exhibits larger scale differences, with evidence of ringing and susceptibility artifacts. \label{fig:alt_data}}
\end{figure}

%% file: main_preprint.bbl
\begin{thebibliography}{51}
\providecommand{\natexlab}[1]{#1}
\providecommand{\url}[1]{\texttt{#1}}
\expandafter\ifx\csname urlstyle\endcsname\relax
  \providecommand{\doi}[1]{doi: #1}\else
  \providecommand{\doi}{doi: \begingroup \urlstyle{rm}\Url}\fi

\bibitem[Alsop et~al.(2015)Alsop, Detre, Golay, G{\"u}nther, Hendrikse,
  Hernandez-Garcia, Lu, MacIntosh, Parkes, Smits, et~al.]{alsop2015recommended}
David~C Alsop, John~A Detre, Xavier Golay, Matthias G{\"u}nther, Jeroen
  Hendrikse, Luis Hernandez-Garcia, Hanzhang Lu, Bradley~J MacIntosh, Laura~M
  Parkes, Marion Smits, et~al.
\newblock Recommended implementation of arterial spin-labeled perfusion mri for
  clinical applications: A consensus of the ismrm perfusion study group and the
  european consortium for asl in dementia.
\newblock \emph{Magnetic resonance in medicine}, 73\penalty0 (1):\penalty0
  102--116, 2015.

\bibitem[An et~al.(2021)An, Zhou, and Ji]{an2021mitochondrial}
Hong An, Bing Zhou, and Xunming Ji.
\newblock Mitochondrial quality control in acute ischemic stroke.
\newblock \emph{Journal of Cerebral Blood Flow \& Metabolism}, 41\penalty0
  (12):\penalty0 3157--3170, 2021.

\bibitem[An and Lin(2003)]{an2003impact}
Hongyu An and Weili Lin.
\newblock Impact of intravascular signal on quantitative measures of cerebral
  oxygen extraction and blood volume under normo-and hypercapnic conditions
  using an asymmetric spin echo approach.
\newblock \emph{Magnetic Resonance in Medicine: An Official Journal of the
  International Society for Magnetic Resonance in Medicine}, 50\penalty0
  (4):\penalty0 708--716, 2003.

\bibitem[An et~al.(2001)An, Lin, Celik, and Lee]{an2001quantitative}
Hongyu An, Weili Lin, Azim Celik, and Yueh~Z Lee.
\newblock Quantitative measurements of cerebral metabolic rate of oxygen
  utilization using mri: a volunteer study.
\newblock \emph{NMR in Biomedicine: An International Journal Devoted to the
  Development and Application of Magnetic Resonance In Vivo}, 14\penalty0
  (7-8):\penalty0 441--447, 2001.

\bibitem[Andersson et~al.(2001)Andersson, Hutton, Ashburner, Turner, and
  Friston]{andersson2001modeling}
Jesper~LR Andersson, Chloe Hutton, John Ashburner, Robert Turner, and Karl
  Friston.
\newblock Modeling geometric deformations in epi time series.
\newblock \emph{Neuroimage}, 13\penalty0 (5):\penalty0 903--919, 2001.

\bibitem[Andreazza et~al.(2010)Andreazza, Shao, Wang, and
  Young]{andreazza2010mitochondrial}
Ana~C Andreazza, Li~Shao, Jun-Feng Wang, and L~Trevor Young.
\newblock Mitochondrial complex i activity and oxidative damage to
  mitochondrial proteins in the prefrontal cortex of patients with bipolar
  disorder.
\newblock \emph{Archives of general psychiatry}, 67\penalty0 (4):\penalty0
  360--368, 2010.

\bibitem[Beal(2003)]{beal2003variational}
Matthew~James Beal.
\newblock \emph{Variational algorithms for approximate Bayesian inference}.
\newblock University of London, University College London (United Kingdom),
  2003.

\bibitem[Berman and Pike(2018)]{berman2018transverse}
Avery~JL Berman and G~Bruce Pike.
\newblock Transverse signal decay under the weak field approximation: Theory
  and validation.
\newblock \emph{Magnetic resonance in medicine}, 80\penalty0 (1):\penalty0
  341--350, 2018.

\bibitem[Blazey et~al.(2019)Blazey, Snyder, Su, Goyal, Lee, Vlassenko,
  Arbel{\'a}ez, and Raichle]{blazey2019quantitative}
Tyler Blazey, Abraham~Z Snyder, Yi~Su, Manu~S Goyal, John~J Lee, Andrei~G
  Vlassenko, Ana~Maria Arbel{\'a}ez, and Marcus~E Raichle.
\newblock Quantitative positron emission tomography reveals regional
  differences in aerobic glycolysis within the human brain.
\newblock \emph{Journal of Cerebral Blood Flow \& Metabolism}, 39\penalty0
  (10):\penalty0 2096--2102, 2019.

\bibitem[Blei et~al.(2017)Blei, Kucukelbir, and McAuliffe]{blei2017variational}
David~M Blei, Alp Kucukelbir, and Jon~D McAuliffe.
\newblock Variational inference: A review for statisticians.
\newblock \emph{Journal of the American statistical Association}, 112\penalty0
  (518):\penalty0 859--877, 2017.

\bibitem[Blockley et~al.(2013)Blockley, Griffeth, Germuska, Bulte, and
  Buxton]{blockley2013analysis}
Nicholas~P Blockley, Valerie~EM Griffeth, Michael~A Germuska, Daniel~P Bulte,
  and Richard~B Buxton.
\newblock An analysis of the use of hyperoxia for measuring venous cerebral
  blood volume: comparison of the existing method with a new analysis approach.
\newblock \emph{Neuroimage}, 72:\penalty0 33--40, 2013.

\bibitem[Chappell et~al.(2008)Chappell, Groves, Whitcher, and
  Woolrich]{chappell2008variational}
Michael~A Chappell, Adrian~R Groves, Brandon Whitcher, and Mark~W Woolrich.
\newblock Variational bayesian inference for a nonlinear forward model.
\newblock \emph{IEEE Transactions on Signal Processing}, 57\penalty0
  (1):\penalty0 223--236, 2008.

\bibitem[Cherukara(2019)]{Cherukara2019thesis}
Matthew~T Cherukara.
\newblock {Non-Invasive Quantification of Cerebral Oxygenation in Ischaemic
  Stroke Using MRI}.
\newblock \emph{DPhil Thesis}, 2019.

\bibitem[Cherukara et~al.(2019)Cherukara, Stone, Chappell, and
  Blockley]{Cherukara2019}
Matthew~T. Cherukara, Alan~J. Stone, Michael~A. Chappell, and Nicholas~P.
  Blockley.
\newblock {Model-based Bayesian inference of brain oxygenation using
  quantitative BOLD}.
\newblock \emph{NeuroImage}, 202, 2019.
\newblock ISSN 10959572.
\newblock \doi{10.1016/j.neuroimage.2019.116106}.

\bibitem[Christen et~al.(2014)Christen, Pannetier, Ni, Qiu, Moseley, Schuff,
  and Zaharchuk]{christen2014mr}
Thomas Christen, NA~Pannetier, Wendy~W Ni, Deqiang Qiu, Michael~E Moseley,
  Norbert Schuff, and Greg Zaharchuk.
\newblock Mr vascular fingerprinting: A new approach to compute cerebral blood
  volume, mean vessel radius, and oxygenation maps in the human brain.
\newblock \emph{Neuroimage}, 89:\penalty0 262--270, 2014.

\bibitem[Coles et~al.(2006)Coles, Fryer, Bradley, Nortje, Smielewski, Rice,
  Clark, Pickard, and Menon]{coles2006intersubject}
Jonathan~P Coles, Tim~D Fryer, Peter~G Bradley, Jurgens Nortje, Peter
  Smielewski, Kenneth Rice, John~C Clark, John~D Pickard, and David~K Menon.
\newblock Intersubject variability and reproducibility of 15o pet studies.
\newblock \emph{Journal of Cerebral Blood Flow \& Metabolism}, 26\penalty0
  (1):\penalty0 48--57, 2006.

\bibitem[Correia and Moreira(2021)]{correia2021oxygen}
S{\'o}nia~C Correia and Paula~I Moreira.
\newblock Oxygen sensing and signaling in alzheimer’s disease: A breathtaking
  story!
\newblock \emph{Cellular and Molecular Neurobiology}, pages 1--19, 2021.

\bibitem[Derdeyn et~al.(2001)Derdeyn, Videen, Grubb, and
  Powers]{derdeyn2001comparison}
Colin~P Derdeyn, Tom~O Videen, Robert~L Grubb, and William~J Powers.
\newblock Comparison of pet oxygen extraction fraction methods for the
  prediction of stroke risk.
\newblock \emph{Journal of Nuclear Medicine}, 42\penalty0 (8):\penalty0
  1195--1197, 2001.

\bibitem[Dillon et~al.(2017)Dillon, Langmore, Tran, Brevdo, Vasudevan, Moore,
  Patton, Alemi, Hoffman, and Saurous]{dillon2017tensorflow}
Joshua~V Dillon, Ian Langmore, Dustin Tran, Eugene Brevdo, Srinivas Vasudevan,
  Dave Moore, Brian Patton, Alex Alemi, Matt Hoffman, and Rif~A Saurous.
\newblock Tensorflow distributions.
\newblock \emph{arXiv preprint arXiv:1711.10604}, 2017.

\bibitem[Diringer et~al.(2000)Diringer, Yundt, Videen, Adams, Zazulia, Deibert,
  Aiyagari, Dacey, Grubb, and Powers]{diringer2000no}
Michael~N Diringer, Kent Yundt, Tom~O Videen, Robert~E Adams, Allyson~R
  Zazulia, Ellen Deibert, Venkatesh Aiyagari, Ralph~G Dacey, Robert~L Grubb,
  and William~J Powers.
\newblock No reduction in cerebral metabolism as a result of early moderate
  hyperventilation following severe traumatic brain injury.
\newblock \emph{Journal of neurosurgery}, 92\penalty0 (1):\penalty0 7--13,
  2000.

\bibitem[Domsch et~al.(2018)Domsch, M{\"u}rle, Weing{\"a}rtner, Zapp, Wenz, and
  Schad]{domsch2018oxygen}
Sebastian Domsch, Bettina M{\"u}rle, Sebastian Weing{\"a}rtner, Jascha Zapp,
  Frederik Wenz, and Lothar~R Schad.
\newblock Oxygen extraction fraction mapping at 3 tesla using an artificial
  neural network: a feasibility study.
\newblock \emph{Magnetic resonance in medicine}, 79\penalty0 (2):\penalty0
  890--899, 2018.

\bibitem[Dunn and Isaacs(2021)]{dunn2021impact}
Jeff~F Dunn and Albert~M Isaacs.
\newblock The impact of hypoxia on blood-brain, blood-csf, and csf-brain
  barriers.
\newblock \emph{Journal of Applied Physiology}, 131\penalty0 (3):\penalty0
  977--985, 2021.

\bibitem[Ernst et~al.(1993)Ernst, Kreis, and Ross]{Ernst1993AbsoluteWater}
T.~Ernst, R.~Kreis, and B.~D. Ross.
\newblock {Absolute Quantitation of Water and Metabolites in the Human Brain.
  I. Compartments and Water}.
\newblock \emph{Journal of Magnetic Resonance, Series B}, 102\penalty0
  (1):\penalty0 1--8, 8 1993.

\bibitem[Fahraeus and Lindqvist(1931)]{fahraeus1931viscosity}
Robin Fahraeus and Torsten Lindqvist.
\newblock The viscosity of the blood in narrow capillary tubes.
\newblock \emph{American Journal of Physiology-Legacy Content}, 96\penalty0
  (3):\penalty0 562--568, 1931.

\bibitem[Friston et~al.(2007)Friston, Mattout, Trujillo-Barreto, Ashburner, and
  Penny]{friston2007variational}
Karl Friston, J{\'e}r{\'e}mie Mattout, Nelson Trujillo-Barreto, John Ashburner,
  and Will Penny.
\newblock Variational free energy and the laplace approximation.
\newblock \emph{Neuroimage}, 34\penalty0 (1):\penalty0 220--234, 2007.

\bibitem[Groves et~al.(2009)Groves, Chappell, and Woolrich]{groves2009combined}
Adrian~R Groves, Michael~A Chappell, and Mark~W Woolrich.
\newblock Combined spatial and non-spatial prior for inference on mri
  time-series.
\newblock \emph{Neuroimage}, 45\penalty0 (3):\penalty0 795--809, 2009.

\bibitem[He et~al.(2016)He, Zhang, Ren, and Sun]{he2016deep}
Kaiming He, Xiangyu Zhang, Shaoqing Ren, and Jian Sun.
\newblock Deep residual learning for image recognition.
\newblock In \emph{Proceedings of the IEEE conference on computer vision and
  pattern recognition}, pages 770--778, 2016.

\bibitem[He and Yablonskiy(2007)]{He2007}
Xiang He and Dmitriy~A. Yablonskiy.
\newblock Quantitative bold: Mapping of human cerebral deoxygenated blood
  volume and oxygen extraction fraction: Default state.
\newblock \emph{Magnetic Resonance in Medicine}, 57\penalty0 (1), 2007.

\bibitem[Henchcliffe and Beal(2008)]{henchcliffe2008mitochondrial}
Claire Henchcliffe and M~Flint Beal.
\newblock Mitochondrial biology and oxidative stress in parkinson disease
  pathogenesis.
\newblock \emph{Nature clinical practice Neurology}, 4\penalty0 (11):\penalty0
  600--609, 2008.

\bibitem[Henriksen et~al.(2021)Henriksen, Gjedde, Vang, Law, Aanerud, and
  Rostrup]{henriksen2021regional}
Otto~M Henriksen, Albert Gjedde, Kim Vang, Ian Law, Joel Aanerud, and Egill
  Rostrup.
\newblock Regional and interindividual relationships between cerebral perfusion
  and oxygen metabolism.
\newblock \emph{Journal of Applied Physiology}, 130\penalty0 (6):\penalty0
  1836--1847, 2021.

\bibitem[Hubertus et~al.(2019)Hubertus, Thomas, Cho, Zhang, Wang, and
  Schad]{hubertus2019using}
Simon Hubertus, Sebastian Thomas, Junghun Cho, Shun Zhang, Yi~Wang, and
  Lothar~Rudi Schad.
\newblock Using an artificial neural network for fast mapping of the oxygen
  extraction fraction with combined qsm and quantitative bold.
\newblock \emph{Magnetic resonance in medicine}, 82\penalty0 (6):\penalty0
  2199--2211, 2019.

\bibitem[Hutchinson et~al.(2002)Hutchinson, Gupta, Fryer, Al-Rawi, Chatfield,
  Coles, O'Connell, Kett-White, Minhas, Aigbirhio,
  et~al.]{hutchinson2002correlation}
Peter~J Hutchinson, Arun~K Gupta, Tim~F Fryer, Pippa~G Al-Rawi, Doris~A
  Chatfield, Jonathan~P Coles, Mark~T O'Connell, Rupert Kett-White, Pawan~S
  Minhas, Franklin~I Aigbirhio, et~al.
\newblock Correlation between cerebral blood flow, substrate delivery, and
  metabolism in head injury: a combined microdialysis and triple oxygen
  positron emission tomography study.
\newblock \emph{Journal of Cerebral Blood Flow \& Metabolism}, 22\penalty0
  (6):\penalty0 735--745, 2002.

\bibitem[Izmailov et~al.(2018)Izmailov, Podoprikhin, Garipov, Vetrov, and
  Wilson]{izmailov2018averaging}
Pavel Izmailov, Dmitrii Podoprikhin, Timur Garipov, Dmitry Vetrov, and
  Andrew~Gordon Wilson.
\newblock Averaging weights leads to wider optima and better generalization.
\newblock \emph{arXiv preprint arXiv:1803.05407}, 2018.

\bibitem[Jenkinson and Smith(2001)]{jenkinson2001global}
Mark Jenkinson and Stephen Smith.
\newblock A global optimisation method for robust affine registration of brain
  images.
\newblock \emph{Medical image analysis}, 5\penalty0 (2):\penalty0 143--156,
  2001.

\bibitem[Jenkinson et~al.(2012)Jenkinson, Beckmann, Behrens, Woolrich, and
  Smith]{jenkinson2012fsl}
Mark Jenkinson, Christian~F Beckmann, Timothy~EJ Behrens, Mark~W Woolrich, and
  Stephen~M Smith.
\newblock Fsl.
\newblock \emph{Neuroimage}, 62\penalty0 (2):\penalty0 782--790, 2012.

\bibitem[Kety et~al.(1948)Kety, Schmidt, et~al.]{kety1948effects}
Seymour~S Kety, Carl~F Schmidt, et~al.
\newblock The effects of altered arterial tensions of carbon dioxide and oxygen
  on cerebral blood flow and cerebral oxygen consumption of normal young men.
\newblock \emph{The Journal of clinical investigation}, 27\penalty0
  (4):\penalty0 484--492, 1948.

\bibitem[Kingma and Welling(2014)]{kingma2013auto}
Diederik~P Kingma and Max Welling.
\newblock Auto-encoding variational bayes.
\newblock \emph{ICLR}, 2014.

\bibitem[Lee et~al.(2018)Lee, Englund, and Wehrli]{lee2018interleaved}
Hyunyeol Lee, Erin~K Englund, and Felix~W Wehrli.
\newblock Interleaved quantitative bold: combining extravascular r2-and
  intravascular r2-measurements for estimation of deoxygenated blood volume and
  hemoglobin oxygen saturation.
\newblock \emph{Neuroimage}, 174:\penalty0 420--431, 2018.

\bibitem[Loshchilov and Hutter(2019)]{loshchilov2018decoupled}
Ilya Loshchilov and Frank Hutter.
\newblock Decoupled weight decay regularization.
\newblock In \emph{International Conference on Learning Representations}, 2019.
\newblock URL \url{https://openreview.net/forum?id=Bkg6RiCqY7}.

\bibitem[Pinna and Colasanti(2021)]{pinna2021neurometabolic}
Antonello Pinna and Alessandro Colasanti.
\newblock The neurometabolic basis of mood instability: The parvalbumin
  interneuron link—a systematic review and meta-analysis.
\newblock \emph{Frontiers in pharmacology}, page 2324, 2021.

\bibitem[Prabakaran et~al.(2004)Prabakaran, Swatton, Ryan, Huffaker, Huang,
  Griffin, Wayland, Freeman, Dudbridge, Lilley,
  et~al.]{prabakaran2004mitochondrial}
S~Prabakaran, JE~Swatton, MM~Ryan, SJ~Huffaker, JT-J Huang, JL~Griffin,
  M~Wayland, T~Freeman, F~Dudbridge, KS~Lilley, et~al.
\newblock Mitochondrial dysfunction in schizophrenia: evidence for compromised
  brain metabolism and oxidative stress.
\newblock \emph{Molecular psychiatry}, 9\penalty0 (7):\penalty0 684--697, 2004.

\bibitem[Raichle et~al.(2001)Raichle, MacLeod, Snyder, Powers, Gusnard, and
  Shulman]{raichle2001default}
Marcus~E Raichle, Ann~Mary MacLeod, Abraham~Z Snyder, William~J Powers, Debra~A
  Gusnard, and Gordon~L Shulman.
\newblock A default mode of brain function.
\newblock \emph{Proceedings of the National Academy of Sciences}, 98\penalty0
  (2):\penalty0 676--682, 2001.

\bibitem[Rezende et~al.(2014)Rezende, Mohamed, and
  Wierstra]{rezende2014stochastic}
Danilo~Jimenez Rezende, Shakir Mohamed, and Daan Wierstra.
\newblock Stochastic backpropagation and approximate inference in deep
  generative models.
\newblock In \emph{International conference on machine learning}, pages
  1278--1286. PMLR, 2014.

\bibitem[Smith(2002)]{smith2002fast}
Stephen~M Smith.
\newblock Fast robust automated brain extraction.
\newblock \emph{Human brain mapping}, 17\penalty0 (3):\penalty0 143--155, 2002.

\bibitem[Sohlin and Schad(2011)]{sohlin2011susceptibility}
Maja~C Sohlin and Lothar~R Schad.
\newblock Susceptibility-related mr signal dephasing under nonstatic
  conditions: Experimental verification and consequences for qbold
  measurements.
\newblock \emph{Journal of Magnetic Resonance Imaging}, 33\penalty0
  (2):\penalty0 417--425, 2011.

\bibitem[Stone and Blockley(2017)]{stone2017streamlined}
Alan~J Stone and Nicholas~P Blockley.
\newblock A streamlined acquisition for mapping baseline brain oxygenation
  using quantitative bold.
\newblock \emph{Neuroimage}, 147:\penalty0 79--88, 2017.

\bibitem[Stone et~al.(2019)Stone, Harston, Carone, Okell, Kennedy, and
  Blockley]{stone2019prospects}
Alan~J Stone, George~WJ Harston, Davide Carone, Thomas~W Okell, James Kennedy,
  and Nicholas~P Blockley.
\newblock Prospects for investigating brain oxygenation in acute stroke:
  Experience with a non-contrast quantitative bold based approach.
\newblock \emph{Human brain mapping}, 40\penalty0 (10):\penalty0 2853--2866,
  2019.

\bibitem[Trapp and Stys(2009)]{trapp2009virtual}
Bruce~D Trapp and Peter~K Stys.
\newblock Virtual hypoxia and chronic necrosis of demyelinated axons in
  multiple sclerosis.
\newblock \emph{The Lancet Neurology}, 8\penalty0 (3):\penalty0 280--291, 2009.

\bibitem[Wismer et~al.(1988)Wismer, Buxton, Rosen, Fisel, Oot, Brady, and
  Davis]{wismer1988susceptibility}
GL~Wismer, RB~Buxton, BR~Rosen, CR~Fisel, RF~Oot, TJ~Brady, and KR~Davis.
\newblock Susceptibility induced mr line broadening: applications to brain iron
  mapping.
\newblock \emph{Journal of computer assisted tomography}, 12\penalty0
  (2):\penalty0 259--265, 1988.

\bibitem[Xu et~al.(2018)Xu, Li, Liu, Hua, Strouse, Pekar, Lu, van Zijl, and
  Qin]{xu2018accounting}
Feng Xu, Wenbo Li, Peiying Liu, Jun Hua, John~J Strouse, James~J Pekar,
  Hanzhang Lu, Peter~CM van Zijl, and Qin Qin.
\newblock Accounting for the role of hematocrit in between-subject variations
  of mri-derived baseline cerebral hemodynamic parameters and functional bold
  responses.
\newblock \emph{Human brain mapping}, 39\penalty0 (1):\penalty0 344--353, 2018.

\bibitem[Yablonskiy and Haacke(1994)]{Yablonskiy1994}
Dmitriy~A. Yablonskiy and E.~Mark Haacke.
\newblock Theory of nmr signal behavior in magnetically inhomogeneous tissues:
  The static dephasing regime.
\newblock \emph{Magnetic Resonance in Medicine}, 32\penalty0 (6), 12 1994.

\end{thebibliography}
